\documentclass[useAMS,usenatbib]{mn2e}
\usepackage{epsfig}

\def\simless{\mathbin{\lower 3pt\hbox
   {$\rlap{\raise 5pt\hbox{$\char'074$}}\mathchar"7218$}}}
\def\simgreat{\mathbin{\lower 3pt\hbox
   {$\rlap{\raise 5pt\hbox{$\char'076$}}\mathchar"7218$}}}

\title[Does $X_{\rm CO}$ depend on the SFR?]{Does the CO-to-H$_2$ conversion factor depend on the star formation rate?}
\author[]{Paul C.~Clark$^{1, 2}$ \& Simon C.~O.~Glover$^{2}$ \\
$^{1}$School of Physics and Astronomy, Queen's Buildings, The Parade, Cardiff University, Cardiff, CF24 3AA \\
$^{2}$Zentrum f\"ur Astronomie der Universit\"at Heidelberg, Institut f\"ur Theoretische
Astrophysik, Albert-Ueberle-Str.\ 2, 69120 Heidelberg \\
 {\tt email:} paul.clark@astro.cf.ac.uk, glover@uni-heidelberg.de}

\begin{document}

\maketitle

\begin{abstract}
We present a series of numerical simulations that explore how the `X-factor', $X_{\rm CO}$  -- the conversion factor between the observed integrated CO emission and the column density of molecular hydrogen -- varies with the environmental conditions in which a molecular cloud is placed. Our investigation is centred around two environmental conditions in particular:  the cosmic ray ionisation rate (CRIR) and the strength of the interstellar radiation field (ISRF). Since both these properties of the interstellar medium have their origins in massive stars, we make the assumption in this paper that both the strength of the ISRF and the CRIR scale linearly with the local star formation rate (SFR). The cloud modelling in this study first involves running numerical simulations that capture the cloud dynamics, as well as the time-dependent chemistry, and ISM heating and cooling. These simulations are then post-processed with a line radiative transfer code to create synthetic $^{12}$CO (1-0) emission maps from which $X_{\rm CO}$ can be calculated. We find that for $10^4 \, \rm M_{\odot}$ virialised clouds with mean density 100 $\rm cm^{-3}$, $X_{\rm CO}$ is only weakly dependent on the local SFR, varying by a factor of a few over two orders of magnitude in SFR. In contrast, we find that for similar clouds but with masses of $10^5 \,\rm M_{\odot}$, the X-factor will vary by an order of magnitude over the same range in SFR, implying that extra-galactic star formation laws should be viewed with caution. However, for denser ($\rm 10^4\, cm^{-3}$), super-virial clouds such as those found at the centre of the Milky Way, the X-factor is once again independent of the local SFR.
\end{abstract}

\begin{keywords}
galaxies: ISM -- ISM: clouds -- ISM: molecules -- molecular processes
\end{keywords}

\section{Introduction}
\label{sec:intro}
In both the Milky Way, and in other local galaxies, star formation is known to take place within large clouds 
of molecular gas, the so-called giant molecular clouds (GMCs). Understanding the properties of these GMCs is
important for the light it sheds on the process of star formation. Unfortunately, the main chemical
constituent of the gas in GMCs, molecular hydrogen (H$_{2}$), is extremely difficult to observe {\em in situ}, 
owing to the fact that the characteristic temperature of the gas in typical GMCs (10--20~K; \citealt{bt07}) is 
much smaller than the temperature required to excite even the lowest rotational transition of the H$_{2}$
molecule. Therefore, most observational studies of GMCs rely on carbon monoxide (CO) as a proxy for molecular
hydrogen. Observations of nearby GMCs find a surprisingly tight correlation between the integrated intensity 
coming from the clouds in the $J = 1\rightarrow 0$ rotational transition line of $^{12}$CO and the molecular
hydrogen column density of the clouds (see e.g.\ \citealt{dick78,sand84,sol87,sm96,dame01}; or the recent
review by \citealt{bol13}), and yield a 
conversion factor between CO intensity\footnote{Unless otherwise noted, when we refer simply to CO from 
this point on, we mean $^{12}$CO.} and H$_{2}$ column density given approximately by \citep{dame01}
\begin{equation}
X_{\rm CO, gal} = \frac{N_{\rm H_{2}}}{W_{\rm CO}} = 2 \times 10^{20} {\rm cm^{-2} \: K^{-1} \: km^{-1} 
\: s},
\end{equation}
where $W_{\rm CO}$ is the velocity-integrated intensity of the CO $J = 1 \rightarrow 0$ 
emission line, averaged over the projected area of the GMC, and $N_{\rm H_{2}}$ is the 
mean H$_{2}$ column density of the GMC, averaged over the same area. This latter quantity cannot be
directly determined from observations of H$_{2}$ emission, but can be inferred from measurements of 
the total column density if the H{\sc i} column density has also been mapped. The total column density
of gas in the cloud can itself be determined for nearby clouds using extinction mapping 
\citep[see e.g.][]{kain09,fr10,pineda10} or measurements of the diffuse $\gamma$-ray flux produced by 
interactions between high energy cosmic rays  and atomic hydrogen, atomic helium and H$_{2}$ 
\citep[e.g.][]{digel99}.

The fact that the values for $X_{\rm CO}$ derived for nearby clouds show very little variation has
led to its widespread use as a general CO-to-H$_{2}$ conversion factor (commonly referred to as
the `X-factor'), even in environments very
different from the Galactic interstellar medium (ISM). It is therefore important to understand when the adoption 
of a constant value for $X_{\rm CO}$ is appropriate, and when it may be seriously misleading. 

In clouds with a fixed H$_{2}$ column density, any variation in $X_{\rm CO}$ that occurs from cloud
to cloud must be due solely to a variation in $W_{\rm CO}$.  Numerical simulations show that this is
sensitive to three main parameters: the velocity dispersion of the gas in the cloud, the temperature of
the CO-emitting gas, and the filling factor of bright CO emission \citep{gm11,Shetty2011a,Shetty2011b,nh13}.
Increasing the gas temperature increases the typical brightness temperature of the emitting gas,
thereby increasing $W_{\rm CO}$ and decreasing $X_{\rm CO}$. However, as a change in the temperature
also affects the excitation of the CO molecules, the scaling of $W_{\rm CO}$ with temperature is not
linear, but instead is closer to $W_{\rm CO} \propto T^{1/2}$ \citep{Shetty2011b}. Increasing the velocity
dispersion also increases $W_{\rm CO}$ by increasing the CO linewidth, but the increased velocity
dispersion can also lead to lower CO line opacities and hence lower brightness temperatures, so again
the dependence is sub-linear \citep{Shetty2011b}. Finally, $W_{\rm CO}$ is sensitive to the fraction of the
cloud traced by bright CO emission, as in cases where this is small, the {\em mean} integrated intensity
is much smaller than the {\em peak} integrated intensity, owing to the effects of beam dilution. 
In solar metallicity clouds situated in a standard Galactic radiation field, the filling factor of the CO
emission is of order unity \citep[see e.g.][or Section 3 below]{wolf10}, and so small variations in its value 
from cloud to cloud have
little effect on $X_{\rm CO}$. At lower metallicities, however, the CO filling factor can become very small
\citep{gm11,gc12b} and hence $X_{\rm CO}$ can increase dramatically.

If we also allow the H$_{2}$ column density to vary, then this adds an additional parameter to the
problem. However, in real clouds, it is unlikely that all of these parameters vary independently.
In particular, we expect the filling factor of CO-bright gas to be sensitive to the mean extinction of
the cloud \citep{gm11}, which at fixed metallicity is directly related to $N_{\rm H_{2}}$. 
Nevertheless, in order to understand why $X_{\rm CO}$ is close to constant in local GMCs, we
need to understand why these four parameters do not all vary by large amounts. 

If GMCs are in virial equilibrium with linewidths that satisfy the standard Galactic size-linewidth 
relationship \citep{sol87,sco87,rd10}
\begin{equation}
\sigma \simeq 0.7 \left(\frac{R}{1 \: {\rm pc}}\right)^{1/2} {\rm km \: s^{-1}},
\end{equation}
where $\sigma$ is the linewidth and $R$ is a measure of the characteristic size of the cloud,
then we would expect that $N_{\rm H_{2}}$ would be approximately constant from cloud to cloud
\citep{larson81} and that the CO linewidth would scale only weakly with the cloud mass as
$\sigma \propto M_{\rm vir}^{1/4}$ \citep{bol13}, thereby explaining much of the constancy in
$X_{\rm CO}$. However, it has become clear over the last few years that this is not the only
possible explanation and that virial equilibrium is not required in order to explain a constant
$X_{\rm CO}$. The reason is that even if GMCs are not virialized, their mean H$_{2}$
column densities are unlikely to vary by a large amount. Clouds with low values of $N_{\rm H_{2}}$
will provide insufficient dust shielding to allow widespread CO formation and hence will not
be identified as ``molecular'' clouds \citep{gm11,cg13}, while clouds with high $N_{\rm H_{2}}$ 
will form stars rapidly, and the resulting stellar feedback will prevent their column densities
from growing too large \citep{feld12,nh13}.

The small variation that we find in the mean CO brightness temperature of local GMCs can
also be easily understood. Since $^{12}$CO emission from GMCs is usually optically thick, the 
observed brightness temperature is determined by the density and temperature of the gas at the 
point in the cloud where $\tau = 1$ \citep{bol13}. These values vary much less from cloud to
cloud than do the mean values for the cloud as a whole \citep{molina11,nh13}, and so there
is little variation in the resulting mean brightness temperatures \citep{rd10,nh13}.

Putting all of this together, we see that we can understand fairly well why $X_{\rm CO}$ is approximately
constant for local GMCs. However, the behaviour of $X_{\rm CO}$ as we move to environments with much
stronger UV radiation fields and higher cosmic ray fluxes is less easy to predict.
We would expect that as we increase the strength of the
ambient radiation field, more photodissociation of CO will occur. This will decrease the filling factor of
the CO emission and may also decrease the CO linewidth if the CO-emitting gas is no longer 
well-distributed throughout the volume of the cloud. These effects will tend to decrease $W_{\rm CO}$ and hence 
increase $X_{\rm CO}$. At the same time, we also expect a stronger UV field to lead to greater dissociation of H$_{2}$, 
and hence a decrease in $N_{\rm H_{2}}$. In addition, we would expect the cloud to be systematically warmer, owing to 
the higher photoelectric heating rate in low $A_{\rm V}$ regions, and the greater cosmic ray heating rate in
high $A_{\rm V}$ regions. This means that the mean brightness temperature of the CO emission may also be
larger. Together, these two effects will tend to decrease $X_{\rm CO}$. It is not immediately clear which set of effects 
will dominate, and hence whether one would expect $X_{\rm CO}$ to increase or decrease as the strength of the 
UV radiation field and the cosmic ray flux increase.

In this paper, we investigate this issue with the help of numerical simulations. We simulate the coupled chemical,
thermal and dynamical evolution of several model clouds in a variety of different environments, spanning a range
of 100 in terms of UV field strength and cosmic ray ionisation rate. We model and analyze the resulting emission
in the 1-0 line of $^{12}$CO, and thereby are able to explore how $X_{\rm CO}$ varies as a function of the
external environment in clouds with a fixed mass and mean volume density. In Section 2, we briefly describe our
numerical approach, and in Section 3, we explore how the changing environment affects the physical structure of
the clouds. In Section 4, we examine how $X_{\rm CO}$ varies from cloud to cloud, and how this depends on the
definition used for the cloud boundary. We discuss some implications of our results in Section 5 and conclude in 
Section 6. 

\begin{figure*}
\centerline{ \includegraphics[width=6.8in]{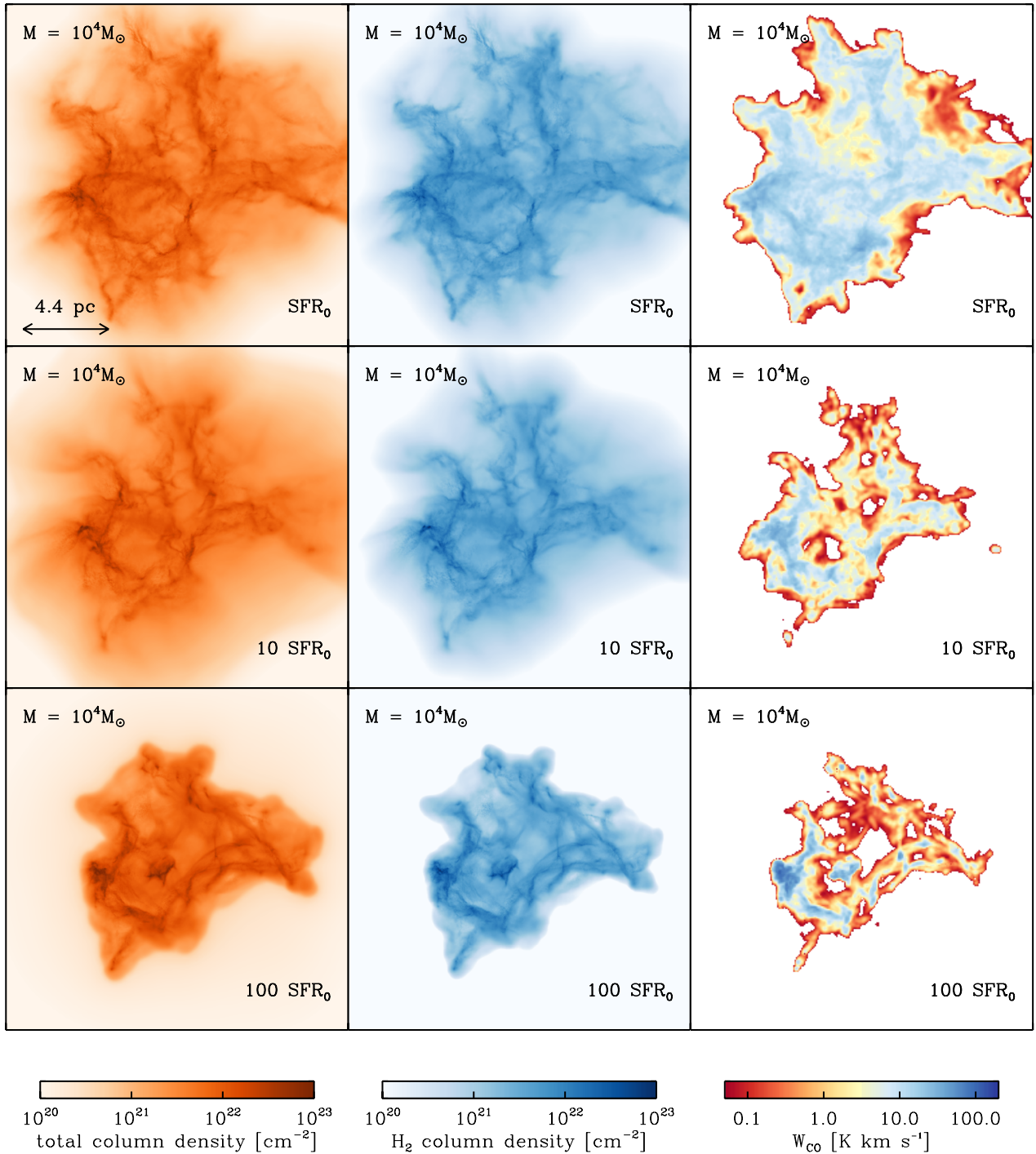} }
\caption{The columns show the total column density (left), the H$_2$ column density (middle) and the velocity-integrated CO intensity in the 1-0 line (right) for the ``standard'' SFR$_0$ (top row), $10 \times$ SFR$_0$ (middle row) and $100 \times$ SFR$_0$ (bottom row) in the $10^4\,\rm M_{\odot}$ clouds. All of the images were made at the point at which the density exceeded $10^6\,\rm cm^{-3}$ in the first collapsing core in each simulation.}
\label{fig:image_m4}
\end{figure*}

\begin{figure*}
\centerline{ \includegraphics[width=6.8in]{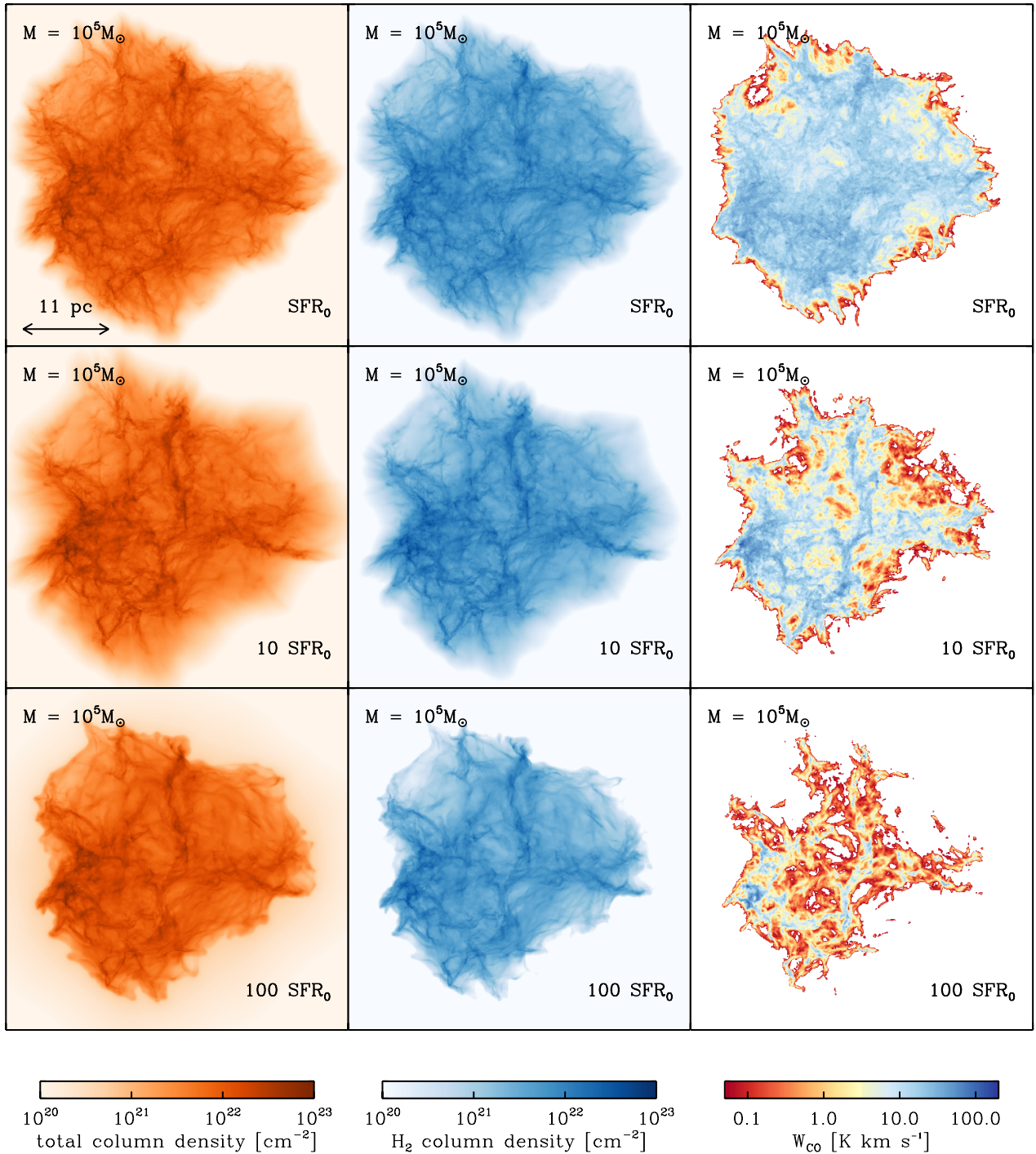} }
\caption{As Figure \ref{fig:image_m4}, but for the $10^5\,\rm M_{\odot}$ clouds. Note the change of physical spatial scale, as illustrated in the top left panel.}
\label{fig:image_m5}
\end{figure*}

\section{Numerical method}
\label{sec:numerics}
\subsection{Details of the algorithms}
The computations presented in this paper were performed using a modified version of the publicly available smoothed particle hydrodynamics (SPH) code {\sc GADGET-2} \citep{springel05}. A major modification to the original {\sc GADGET-2} is the inclusion of time-dependent chemistry. The version we use here follows the formation and destruction of H$_2$ as introduced by \citet{gm07a,gm07b} in addition to the simplified treatment of the CO chemistry that was proposed by \citet{nl99}. In \citet{gc12a}, it was shown that this CO network has a similar accuracy to the more exhaustive treatment of \citet{glo10}, but incurs around one third of the computational cost. We do not include freeze-out of CO onto dust grains in this model, but as this generally occurs only in regions where the $^{12}$CO emission is already optically thick \citep{gold01}, we do not expect this omission to significantly affect our results.

As well as following the chemical evolution of the gas, we also model its thermal evolution. We account for dynamical heating due to shocks and adiabatic compression 
and cooling due to adiabatic rarefaction in the same fashion as in the unmodified version of {\sc GADGET-2}. In addition, we also account for the main radiative and chemical heating and cooling processes occurring in the ISM. These include fine structure cooling from C$^{+}$, C and O, molecular line cooling from H$_{2}$ and CO, photoelectric heating and cosmic ray heating. At high gas densities, collisions between gas particles and dust grains also play an important role in regulating the thermal energy balance, cooling the gas if $T_{\rm gas} > T_{\rm dust}$ and heating it if $T_{\rm dust} > T_{\rm gas}$. Full details of how we treat these processes and a number of other, less important, contributors to the overall thermal energy balance can be found in our previous papers \citep{gm07a,gm07b,glo10,gc12a}, and a summary of the most important processes included can be found in Figure 4 in \citet{clark13}.


The attenuation of the interstellar radiation field (ISRF) is treated using the {\sc TreeCol} algorithm, introduced by \citet{cgk12}. In this paper, the spectral shape of the ISRF is based on the prescription of \citet{dr78} in the ultraviolet and \citet{bl94} at longer wavelengths. The strength of the ISRF is varied in the different simulations as described in Section~\ref{sec:sfr-proxy} below. The clouds are assumed to be bathed in a uniform ISRF, and {\sc TreeCol} is used to compute the attenuated spectrum that reaches each SPH particle in the computational volume. 

Since the above-mentioned papers, we have made two significant changes to the chemical model. The first was to update our treatment of the photodissociation of CO from the prescription given in \citet{lee96}, to that described in the recent paper of \citet{visser09}. The second change that we have made to the chemical model is the inclusion of the effects of  cosmic-ray ionisation of atomic carbon
\begin{equation}
{\rm C} + {\rm c. r.} \rightarrow {\rm C^{+}} + {\rm e^{-}}, \label{react1}
\end{equation}
and cosmic-ray induced photodissociation of C and CO
\begin{eqnarray}
{\rm C} + \gamma_{\rm cr} & \rightarrow &  {\rm C^{+}} + {\rm e^{-}},  \label{react2} \\
{\rm CO} + \gamma_{\rm cr} & \rightarrow &  {\rm C + O}. \label{react3}
\end{eqnarray}
These processes were not included in the chemical model described in \citet{gc12a}. We assume that the rates of all three processes are proportional to $\zeta_{\rm H}$, the cosmic ray ionisation rate of atomic hydrogen. For processes~\ref{react1} and \ref{react2}, we use the rates given in \citet{umist07} as a basis, but rescale them by a factor $\zeta_{\rm H} / \zeta_{\rm H, W07}$, where $\zeta_{\rm H, W07}$ is the value of the cosmic ray ionisation rate adopted by \citet{umist07}. Similarly, to compute the rate of process~\ref{react3}, we use the value given in \citet{gld87} as a basis, but rescale it to make it consistent with our choice of $\zeta_{\rm H}$.

The $^{12}$CO (1-0) emission maps that form the basis of the analysis in this paper were created using the {\sc RADMC-3D} radiative transfer code\footnote{http://www.ita.uni-heidelberg.de/$\sim$dullemond/software/radmc-3d/}. The level populations in the non-LTE limit are computed using the the large velocity gradient approximation \citep{Sobolev57}, as implemented in {\sc RADMC-3D} by \citet{Shetty2011a, Shetty2011b}. We assume that collisions with H$_{2}$ dominate the excitation of the CO rotational levels and adopt the values for the collisional excitation rate coefficients given in the Leiden Atomic and Molecular database \citep{sch05}.

\subsection{Cloud properties}

All of the clouds in this study start as uniform spheres of gas, onto which a three-dimensional turbulent velocity field has been superimposed. Two cloud masses are examined in this paper: a ``low-mass'' cloud, with a mass of $10^4\,\rm M_{\odot}$, and a ``high-mass'' cloud of mass $10^5\,\rm M_{\odot}$. In most of the runs, we take the initial hydrogen nuclei number density of the gas to be $n_{0} = 100 \: {\rm cm^{-3}}$, yielding initial radii for the low- and high-mass clouds of 8.8~pc and 19~pc, respectively. In addition, we performed two simulations of high-mass clouds with a much higher initial density, $n_{0} = 10^{4} \: {\rm cm^{-3}}$; these denser clouds had an initial radius of around 4~pc.

The mass resolution in this study is kept fixed, with the mass of an SPH particle being set at 0.005$\,\rm M_{\odot}$. The low-mass clouds therefore have $2 \times 10^6$ SPH particles,  while the high-mass clouds have $2 \times 10^7$ SPH particles. The minimum resolvable self-gravitating mass-element in this calculation is therefore 0.5$\,\rm M_{\odot}$ in both cases \citep{bb97}. In practice, this means that we can follow the collapse of the cloud until the gas number density reaches a value of around $10^6 \, \rm cm^{-3}$. At this point, we halt the collapse and perform the analysis that is presented in this paper. We note that we would not expect the results presented here to change if we were to follow the collapse to higher densities, as at these densities, the CO (1-0) line will be highly optically thick \citep[see e.g.][]{gold01}, and in any case much of the CO will be frozen out onto dust grains.

The velocity fields are generated with a `natural' mix of solenoidal and compressive modes (i.e.\ a ratio of 2:1). This is generated on a 128$^3$ grid. This velocity field is left to freely decay in shocks, rather than being continuously driven during the course of the simulation. In most of the simulations, the initial energy in the velocity field, $E_{\rm kin}$, was set to be half the initial gravitational potential energy of the cloud, $E_{\rm grav}$, so that the clouds are initially in virial equilibrium. This means that the initial RMS turbulent velocity $v_{\rm rms}$ is 2.4 km\,s$^{-1}$ for the low-mass clouds and 5.2  km\,s$^{-1}$ for the high-mass clouds. In terms of the virial parameter
\begin{equation}
\label{eq:alphavir}
\alpha_{\rm vir} = \frac{E_{\rm kin}}{E_{\rm grav}},
\end{equation}
these clouds have $\alpha_{\rm vir} = 0.5$. In addition, we also examine one high density cloud for which we set $\alpha_{\rm vir} = 2$, so that the cloud is initially unbound (see Section~\ref{Sect:SFR}).  Note that most of the simulations presented here adopt the same random seed for the turbulent velocity field. However we also run one of the simulations with a different seed, to gauge the sensitivity of our results to the underlying cloud structure that is created by the turbulent motions.

From previous modelling, it has been found that in cold gas with densities of around $100\,\rm cm^{-3}$, most of the hydrogen is in the form of H$_2$. In contrast, the carbon in this gas is still predominantly in the form of C$^+$ \citep[see e.g.][]{clark12,smith14}. The initial chemical state of the gas in the majority of the runs presented in this study is motivated by this previous work. We start most of our runs with all of the hydrogen already in the form of H$_{2}$, but assume that the carbon and oxygen are present in the form of C$^{+}$ and O, respectively. We adopt total abundances (relative to the number density of hydrogen nuclei) for the carbon and oxygen nuclei of $x_{\rm C} = 1.4 \times 10^{-4}$ and $x_{\rm O} = 3.2 \times 10^{-4}$ respectively, consistent with the values measured in the local ISM \citep{sembach00}. Similarly, we adopt a dust-to-gas mass ratio of 0.01, consistent with the value in the local ISM.

In our runs with very high UV fields, it is unclear whether starting with all of the hydrogen in molecular form is a good approximation, as in this case, one would expect the equilibrium H$_{2}$ fraction in low density gas to be much smaller than in the models of \citet{clark12} and \citet{smith14}. To address this uncertainty, we ran two additional high-mass models
with the hydrogen initially in atomic form.

The post-processing of the simulation data in {\sc RADMC-3D} first requires that the SPH particle data is interpolated onto a regular Cartesian grid. This is done using the standard SPH smoothing formalism. To ensure that we catch small, high density pockets of gas, we employ a {\sc RADMC-3D} cell-size of 0.068\,pc, such that the high-mass cloud calculations have $656^3$ cells and the low-mass cloud calculations have $256^3$ cells. We find that such a resolution is sufficient to get converged values for the probability density function of CO (1-0) emission, from which it follows that the various mean values that we examine later are also converged. Once the position-position-velocity maps are obtained, we then integrate the emission along the z-axis to create maps of the integrated intensity, $W_{\rm CO (1-0)}$.

\subsection{The ``star formation rate''}
\label{sec:sfr-proxy}

In this study, we vary two of the environmental conditions that can affect the chemical balance of clouds: the strength of the ISRF and the cosmic-ray ionisation rate. Both of these are thought to vary with the local star formation rate, and so we assume here that the strength of these processes can be used as a proxy for the star formation rate.

As mentioned above, the ISRF used here is taken to have a shape described by a combination of the \citet{bl94}  and \citet{dr78} radiation fields. In one set of runs -- those representing clouds in an environment with a star formation rate similar to that in the local ISM, which we denote as SFR$_{0}$ -- we adopt the same normalization for the ISRF as in the papers of \citet{dr78} and \citet{bl94}. Our fiducial ISRF therefore has a strength $G_{0} = 1.7$ in 
\citet{habing68} units.

For this study, we are mainly interested in how the ISRF heats the gas and affects its chemical state. Photons with energies above 6~eV are responsible for the photoelectric heating (the dominant heat source in low extinction regions of these clouds), and photons with energies above 11.2~eV and 11.5~eV are responsible for dissociating H$_2$ and CO respectively. Since most of the photons in this part of the ISRF come from massive, young stars, it is reasonable to assume that, to a first approximation, the strength of the relevant portion of the ISRF scales linearly with the local star formation rate. At longer wavelengths, the ISRF is dominated by older stellar populations and this assumption is less well-founded. However, the strength of the ISRF at these wavelengths has little effect on the temperature or chemistry of the gas, and so for simplicity, we assume that in regions with higher star formation rates, we can simply scale the entire ISRF upwards, rather than changing its spectral shape. 

We also assume that the cosmic ray ionisation rate scales linearly with the star formation rate. This assumption is reasonable: supernova remnants are the main source of cosmic rays in the Galactic ISM, and the lifetime of a typical cosmic ray within the Galaxy is around 15~Myr \citep{fer01}, so the cosmic ray energy density, and hence the cosmic ray ionisation rate, should track the star formation rate fairly closely.

In this study, in addition to the runs representing the behaviour of clouds in the local ISM, with ${\rm SFR = SFR_{0}}$, we also perform simulations 
where we increase the strength of the ISRF and the cosmic ray ionisation rate by factors of 10 or 100, corresponding to star formation rates of 
$10 \, {\rm SFR_{0}}$ or $100 \, {\rm SFR_{0}}$, respectively.

\begin{table*}
\caption{List of simulations. All number densities are given with respect to the number of hydrogen nuclei, since this is invariant of the chemical state of the gas. The definition of $\alpha_{\rm vir}$ can be found in Equation \ref{eq:alphavir}, while SFR$_{0}$ is defined in Section~\ref{sec:sfr-proxy}. The property  $t_{\rm end}$ denotes the point at which the simulation is stopped, and the cloud details are fed to the line radiative transfer code. This is also the time at which the first star forms in the simulation.  \label{tab:sims}}
\begin{tabular}{cccccc}
\hline
Mass (${\rm M}_{\odot}$) & $n_{0}$ (${\rm cm^{-3}}$) & $\alpha_{\rm vir}$ & SFR (SFR$_{0}$) & Notes & $t_{\rm end}$ (Myr) \\
\hline
$10^{4}$ & 100 & 0.5 & 1 &  & 1.83 \\
$10^{4}$ & 100 & 0.5 & 10 &  & 2.09 \\
$10^{4}$ & 100 & 0.5 & 100 & & 1.91 \\
$10^{4}$ & 100 & 0.5 & 100 & Different seed & 2.18 \\
$10^{5}$ & 100 & 0.5 & 1 &  & 1.17\\
$10^{5}$ & 100 & 0.5 & 10 &  & 1.52 \\
$10^{5}$ & 100 & 0.5 & 100 &  & 1.39 \\
$10^{5}$ & 100 & 0.5 & 1 & Atomic ICs & 1.31 \\
$10^{5}$ & 100 & 0.5 & 100 & Atomic ICs & 1.26 \\
$10^{5}$ & $10^{4}$ & 0.5 & 100 & Galactic Centre style cloud & 0.1 \\
$10^{5}$ & $10^{4}$ & 2 & 100 & Galactic Centre style cloud & 0.1 \\
\hline
\end{tabular}
\end{table*}

\section{General cloud structure}

In the left-hand columns of Figures \ref{fig:image_m4} and \ref{fig:image_m5}, we show column density images of the gas, taken at a point in the simulations just before the onset of star formation in our fiducial clouds -- i.e.\ those seeded with our standard turbulent velocity field, and which started life with all their hydrogen in the form of H$_2$. This particular stage in the cloud's evolution is when we perform the analysis that forms the basis of this paper. For most of our simulated clouds, this corresponds to a time $t \sim 1$--2~Myr after the beginning of the simulation, as shown in Table~\ref{tab:sims}.

\begin{figure}
\centerline{ \includegraphics[width=3.4in]{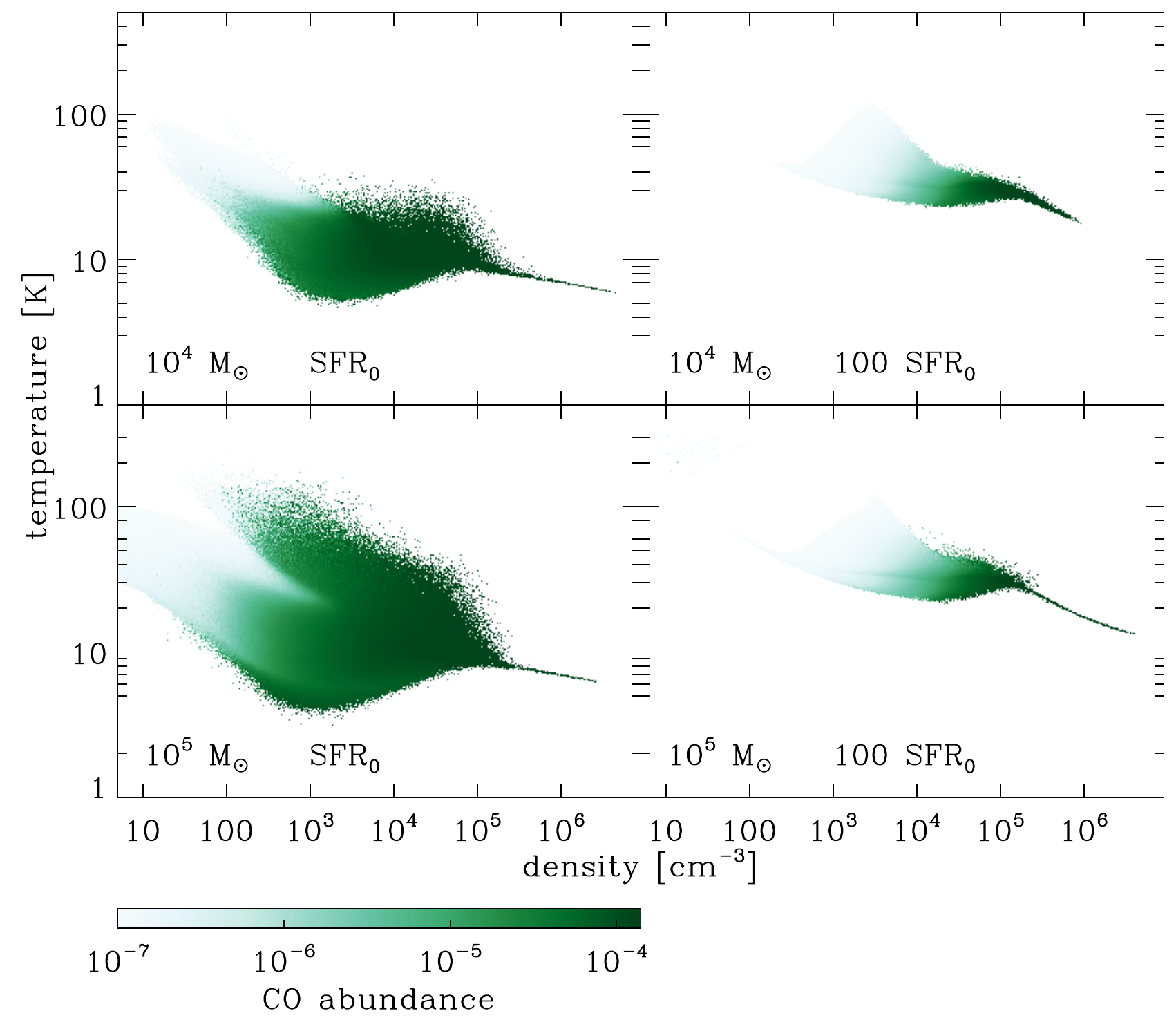} }
\caption{Phase diagrams for four of our simulated clouds, illustrating the temperature structure of the gas at the end of the simulation. The diagrams are colour-coded using the mean CO abundance at each point in temperature-density phase space.}
\label{fig:phase}
\end{figure}

In both the $10^4\,\rm M_{\odot}$ and $10^5\,\rm M_{\odot}$ clouds we see a similar change in behaviour of the strength of the ISRF and cosmic ray ionisation rate are increased to mimic progressively higher rates of ambient star formation activity: the filamentary structure imposed by the cloud's turbulence and self-gravity is less pronounced as we move to higher SFRs. This is simply a result of gas temperatures rising due to the combined effects of the enhanced photoelectric emission and cosmic-ray ionisation rates. This systematic increase in the temperature as we increase the SFR is also illustrated in Figure~\ref{fig:phase}, where we show temperature-density phase diagrams for four of our runs, colour-coded by the CO abundance.

In addition, we see from Figures \ref{fig:image_m4} and \ref{fig:image_m5} that the overall morphology of the clouds changes as the SFR increases. In this respect, there are two trends. The first is that the dense, filamentary structure becomes confined to a progressively smaller region as the SFR increases. This makes sense, as the gas is more easily structured near the centre of the cloud where the extinction is higher and the heating effects of the photoelectric emission are less. As the SFR increases, the heating by photoelectric emission becomes progressively more important in shaping the gas, and so the regions structured by the turbulence and gravity are those towards the centre. 

The second effect is that the structure of the cloud envelope changes. The SPH particles have open (i.e.\ vacuum) boundary conditions, and so the particles are free to expand into the surrounding space. The stronger the ISRF, the more the surface of the cloud is heated, causing it to ``boil off'' into the void, creating the halo that we can see around the clouds. In the case of the 100 $\rm SFR_0$ clouds (i.e.\ those illuminated by radiation fields and cosmic ray fluxes that are 100 times greater than the values in the local ISM; see Section~\ref{sec:sfr-proxy}), the ISRF is so strong that this halo becomes hot enough to push back on the cloud, creating the sharp boundary that we see in Figures \ref{fig:image_m4} and \ref{fig:image_m5}. Although dramatic, this feature of the increasing SFR is actually of little interest to this current study, as both the H$_2$ and CO lie within the cloud boundary.

The middle panels in Figures \ref{fig:image_m4} and \ref{fig:image_m5} show the H$_2$ column densities images at the same point in the cloud evolution. In general, we see that the H$_2$ column density tracks the total column density extremely well for the bulk of the cloud. Given that all the clouds in these two figures were initialised with all their hydrogen in molecular form, this is not surprising. 

The main difference between the left and middle panels in Figures \ref{fig:image_m4} and \ref{fig:image_m5} is that the H$_2$ in the outer envelope has been photodissociated by the ISRF, and so does not trace the very low density (both column and volume) that surrounds the main cloud structure. We also see that the column density at which H$_2$ starts to appear increases with increasing SFR, moving from around $10^{20}\,\rm cm^{-2}$ to around $10^{21}\,\rm cm^{-2}$ as the SFR increases from the solar neighbourhood value to a factor of 100 greater.  This behaviour is consistent with the predictions of detailed models for the structure of photodissociation regions \citep[see e.g.][]{krum08,stern14}. The fact that a large change in the strength of the ISRF leads to a relatively small change in the column density of gas required to effectively shield H$_{2}$ is a consequence of the fact that dust shielding is an exponential function of the column density, which implies that in the dust-dominated region, the critical column density scales only logarithmically with the SFR.

The integrated emission in the 1-0 line of $^{12}$CO, $W_{\rm CO}$, is shown in the right-hand column in Figures \ref{fig:image_m4} and \ref{fig:image_m5}, and shows the most striking variation with increasing SFR.
For the SFR$_0$ clouds, we see that the surface filling factor of the bright CO emission is very similar to that of the  H$_{2}$ gas. Despite this, it is not a good tracer of the physical structure of the cloud: some features seen in the emission map are far more blobby than their H$_{2}$ counterparts and others are missing entirely. This behaviour is expected \citep[see e.g.][]{bpm02,good09,Shetty2011a, Shetty2011b,beau13} and is simply a result of the large optical depth of the 1-0 line -- the features that we see are the $\tau \sim 1$ surfaces, which are a function of CO column density, velocity, and temperature, and so often bear little resemblance to the structure we see in the H$_2$ column density images. 

As we move to higher SFRs, we see that the surface filling factor of the CO emission progressively decreases, a feature we see in all of the cloud models run in this study. In the case of the $100\,\rm SFR_0$ clouds, the effect is extreme, with emission falling below $1\,\rm K\,km\,s^{-1}$ for much of the cloud's interior -- the sensitivity limit in many large-scale surveys -- and even below $0.1\,\rm K\,km\,s^{-1}$ along lines of sight with H$_2$ column densities as high as a few $10^{21}\,\rm cm^{-2}$.

However, in the cases with the higher SFR, we see that the {\em peak} CO emission becomes brighter towards regions of higher column density. This is a consequence of increased heating in the cloud, which results in a higher overall excitation temperature of the CO.
The broader line-width in the gas also helps to the lower the optical depth, allowing more of the emission from the line to escape \citep{Shetty2011a, Shetty2011b}. Due to the higher spatial resolution in the reproduced image (the true resolution is the same in all images), this effect is perhaps most easily seen in the $10^4\,\rm M_{\odot}$ clouds in Figure \ref{fig:image_m4}. Here we see the brightening of the clump towards the left of the $W_{\rm CO(1-0)}$ image, which goes from having a peak brightness of around $30\,\rm K\,km\,s^{-1}$ for SFR$_0$, to nearly $80\,\rm K\,km\,s^{-1}$ in the case of 100\,SFR$_{0}$

\section{X-factor variations}

\subsection{Computing the X-factor}

\begin{figure}
\centerline{ \includegraphics[width=3.30in]{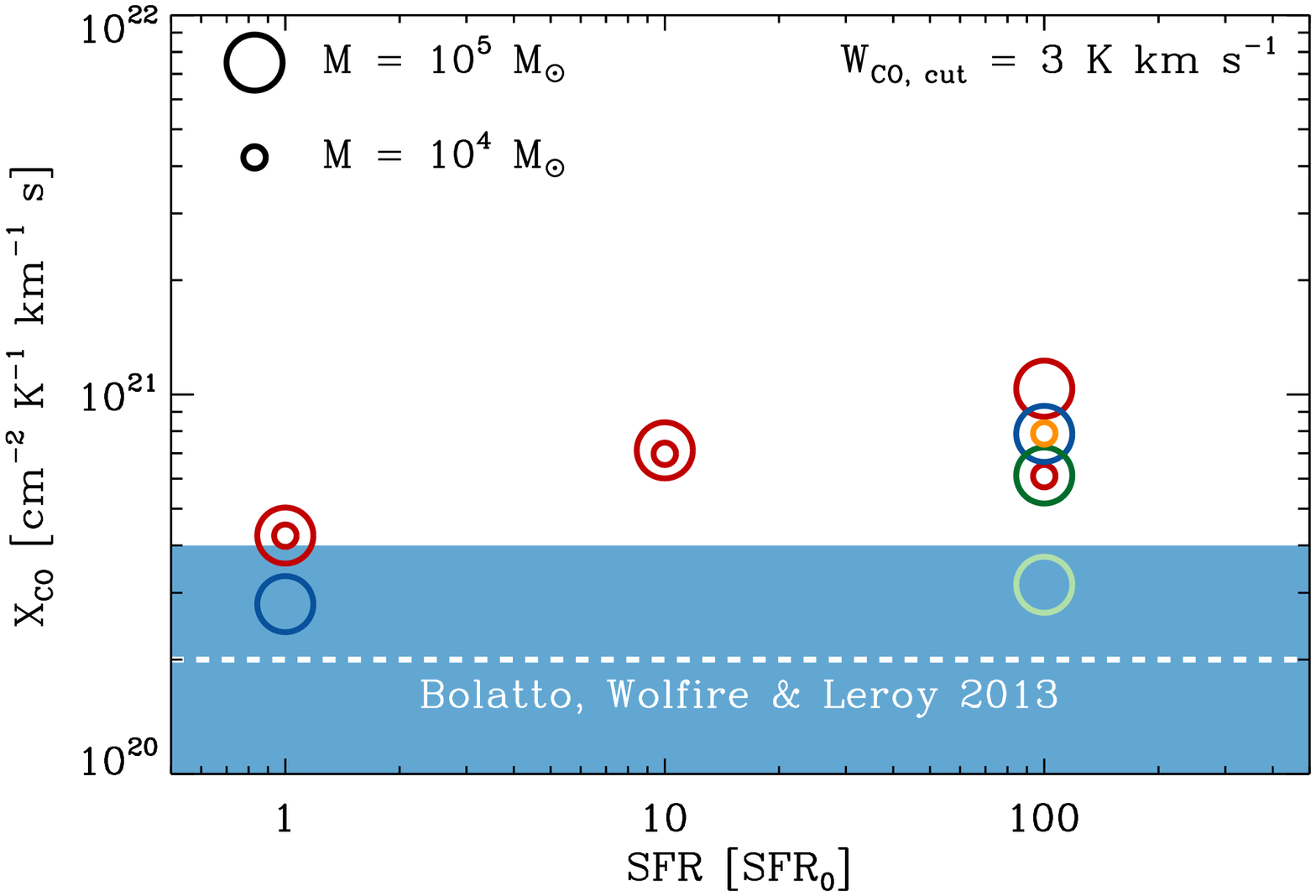} }
\centerline{ \includegraphics[width=3.30in]{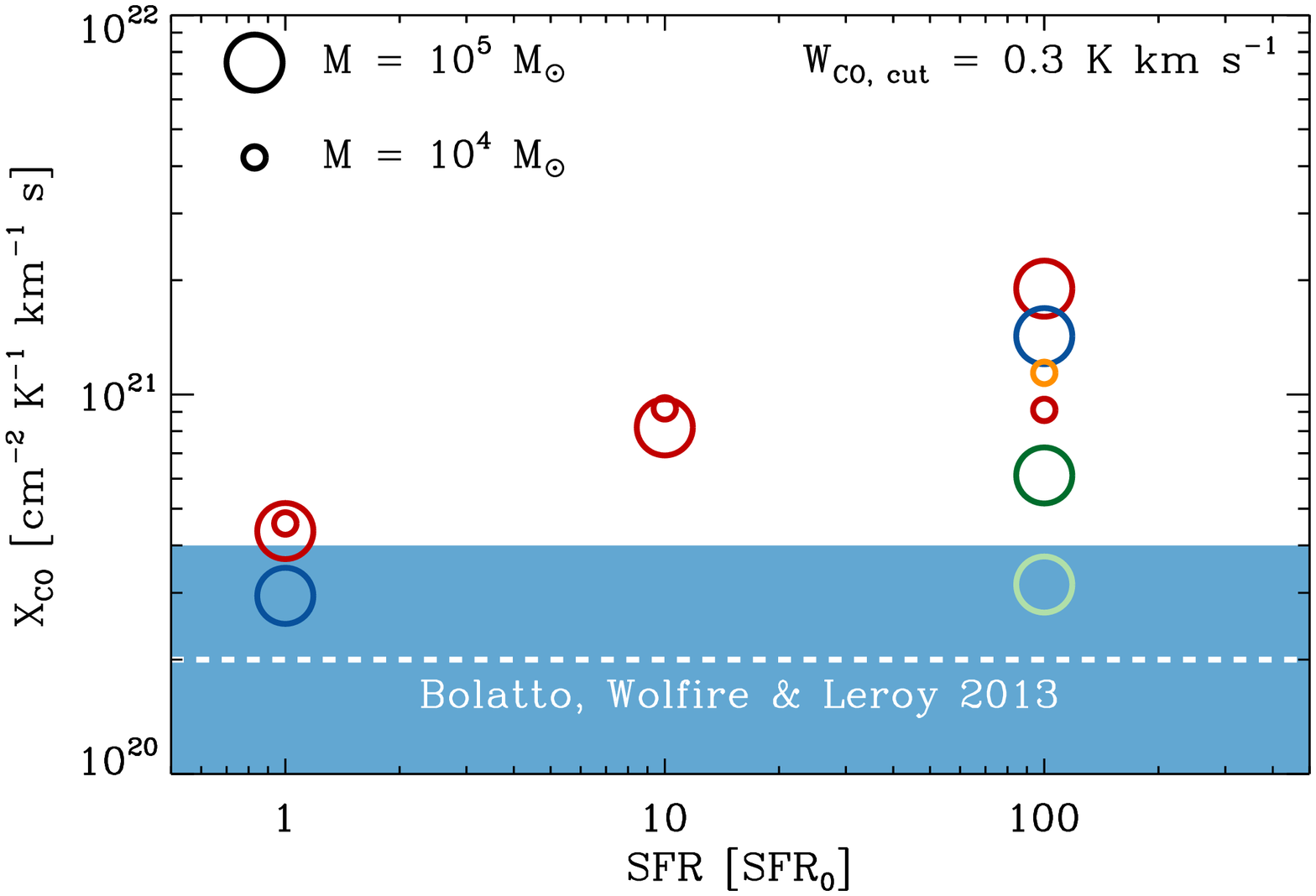} }
\centerline{ \includegraphics[width=3.30in]{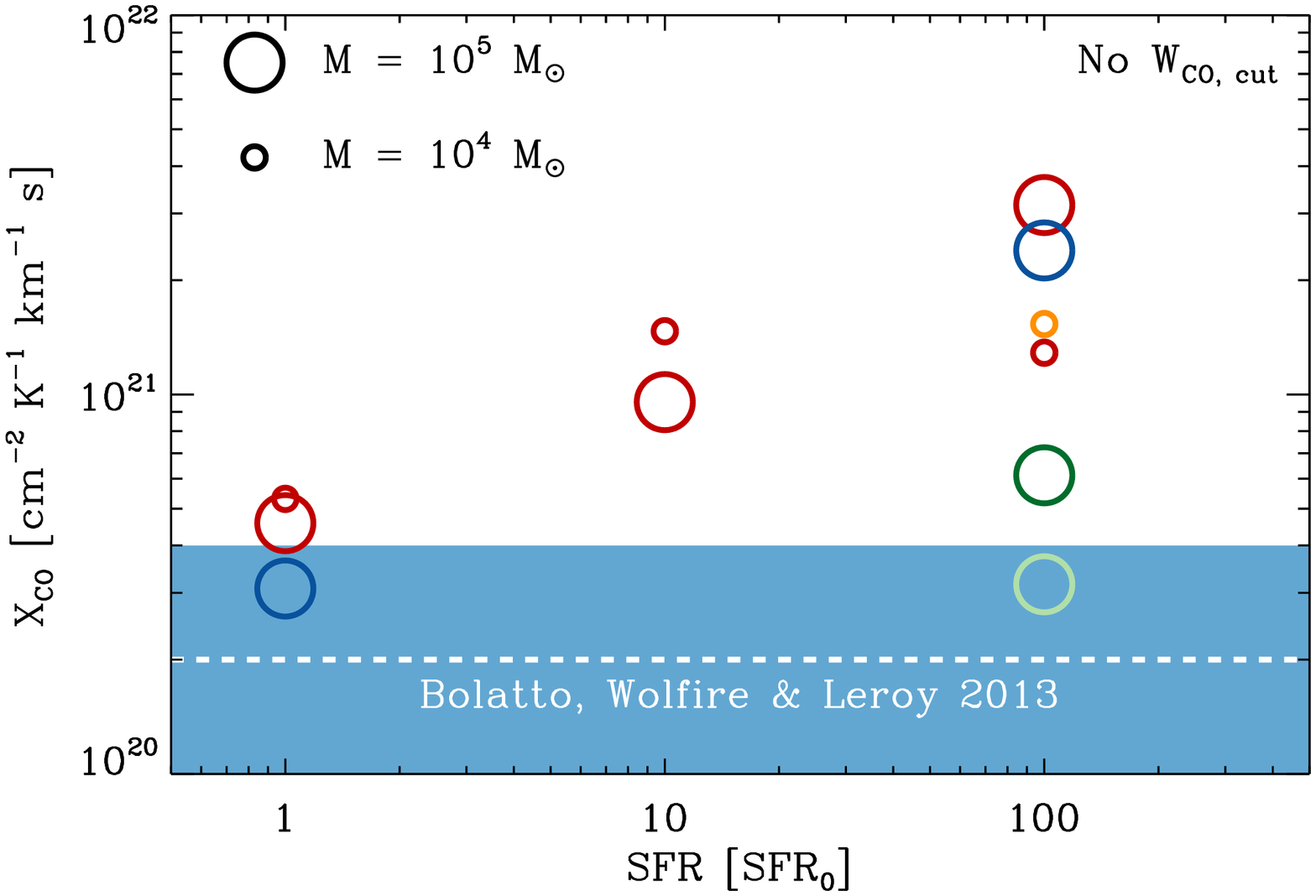} }
\caption{Mean CO-to-H$_{2}$ conversion factor, $X_{\rm CO}$, in each of our clouds, as a function of our SFR proxy (see Section \ref{sec:sfr-proxy} for details).
Red and blue circles denote the results from the $n_0 = 100 \, \rm cm^{-3}$ clouds with fully molecular and fully atomic initial conditions respectively. 
The orange circles correspond to a cloud with $n_0 = 100 \, \rm cm^{-3}$ run using a different random seed for the turbulent velocity field.
The dark and light green circles denote the $n_0 = 10^4 \, \rm cm^{-3}$ clouds with, respectively, $\alpha_{\rm vir} = 0.5$ and 2. 
In the top two panels, we compute $X_{\rm CO}$ using integrated intensities and H$_{2}$ column densities only for those lines-of-sight with an integrated intensity above the specified threshold, $W_{\rm CO, cut}$. In the bottom panel, we show the results that we obtain if we do not apply an integrated intensity threshold. In each case, we compute $X_{\rm CO}$ by dividing the mean H$_{2}$ column density in the considered area by the mean value of $W_{\rm CO}$ from the same area. The horizontal dashed line shows the canonical local value of $X_{\rm CO}$, and the shaded area indicates the typical scatter in the value of  $X_{\rm CO}$ derived observationally for the Milky Way and other nearby spiral galaxies \citep[see][]{bol13}.}
\label{fig:xco_sfr}
\end{figure}

In the previous section, we saw that as the SFR in the ambient cloud environment is increased, the CO emission traces progressively less of the H$_2$ distribution. However we have also seen that the peak emission in the cloud is higher in the case of high SFR. \citet{bol13} have suggested that these two effects may effectively cancel out, resulting in an X-factor that does not depend on the cloud's environment. In this scenario, the greater emission that comes from the dense bright peaks in the high SFR case makes up for the fact that much of the cloud is not emitting in CO at all, since the molecule is easily photodissociated. Using our self-consistent models for N$_{\rm H_{2}}$ and $W_{\rm CO}$, we can now explore this idea.

In what follows, we compare how $X_{\rm CO}$ varies as we alter the environmental conditions in our suite of turbulent clouds. We calculate $X_{\rm CO}$ for each cloud via
\begin{equation}
\label{eq:xcopix}
X_{\rm CO} = \frac{ \langle N_{\rm H_2} \rangle } { \langle W_{\rm CO} \rangle }
\end{equation}
where the averages are taken over a specified set of the pixels in the images from Figures \ref{fig:image_m4} and \ref{fig:image_m5}.

In our evaluation of Equation \ref{eq:xcopix}, we take two approaches. First, we examine the values of X$_{\rm CO}$ obtained when we average over all of the pixels in the images. In our second approach, we instead restrict the calculation to include only those pixels with CO 1-0 integrated intensities above some threshold value, $W_{\rm CO, cut}$. These two approaches mimic different types of observations. The first approach is more appropriate when considering unresolved molecular clouds that could inhabit the beam in an extragalactic CO observation. In this case, the CO emission may not trace all of the H$_2$ gas in the beam, i.e.\ there is a `dark' molecular component surrounding the CO-bright portion of the cloud, as recently highlighted by \citet{wolf10}. The second approach, in which we only examine pixels with emission above $W_{\rm CO, cut}$, is more applicable to studies of nearby, resolved molecular clouds, where it is common to define the extent of a cloud in terms of its CO emission, a definition which inevitably leads to one ignoring those regions without detectable CO emission \citep[see e.g.][]{pineda08,myl14}.

The mean values of $X_{\rm CO}$ that we derive from the simulations, and how they scale with the SFR, are shown in Figure \ref{fig:xco_sfr}. The top two panels show the results that we obtain for two different CO detection thresholds, $W_{\rm CO, cut} = 3 \: {\rm K \, km \, s^{-1}}$ and $W_{\rm CO, cut} = 0.3 \: {\rm K \, km \, s^{-1}}$, while the bottom panel shows the results in the case where we do not apply a detection threshold. We also indicate the canonical local value of the X-factor, $X_{\rm CO, gal}$ (the horizontal dashed  line) and illustrate the typical scatter in the value of $X_{\rm CO}$ derived for the disks of nearby spiral galaxies (the shaded region), as summarized in \citet{bol13}.

\subsection{Clouds with SFR = SFR$_{0}$}
\label{sfr0}
We begin our study of the effects of the local environment on $X_{\rm CO}$ by looking at the behaviour of clouds located in an environment with a star formation rate similar to the local ISM, i.e.\ with ${\rm SFR = SFR_{0}}$ (see Section~\ref{sec:sfr-proxy}). The first point to note is that when we consider clouds in this environment that have properties similar to those of local molecular clouds (i.e.\ virial parameters close to unity and mean initial densities around 100 $\rm cm^{-3}$), we find that
$X_{\rm CO}$ does not appear to be sensitive to mass. Both our low-mass and our high-mass clouds yield a mean X-factor of between 4-5 $\times 10^{20} \,{\rm cm^{-2} K^{-1} \, km^{-1} \,s}$, consistent with the large body of literature suggesting that the X-factor is not strongly dependent on the cloud mass \citep[see e.g.][]{sol87,blitz07,hughes10,bol13}. Also, we find that the X-factor is relatively insensitive to the value of  $W_{\rm CO, cut}$, indicating that the CO emission is a good proxy for the total H$_2$ present in such clouds, even though the emission is clearly not tracing all of the molecular gas. 

While this is encouraging, it should also be noted that the values we recover for $X_{\rm CO}$ in this case are roughly twice as large as the canonical Galactic value, $X_{\rm CO, gal}$, and on the edge of the range of values found for the disks of nearby spiral galaxies \citep{bol13}. The reason for this 
discrepancy is not completely clear, but there are several factors that may contribute to our recovering a systematically higher $X_{\rm CO}$. 
The first of these stems from the way in which we treat the effects of self-shielding in our simulations. 
When calculating each particle's sky maps of H$_2$ and CO column densities, the {\sc TreeCol} algorithm includes contributions from all the molecules present in the cloud, regardless of the velocity at which they are moving with respect to the target particle. In reality, much of the cloud will not contribute to the self-shielding of a given location, since the cloud's supersonic motions will Doppler shift the molecules out of the line-profile. Neglecting this effect means that we overestimate the effectiveness of H$_{2}$ and CO self-shielding, and hence overproduce the abundances of both molecules. We expect the error to be larger for H$_{2}$ than for CO, since self-shielding plays a much greater role in determining the H$_{2}$ abundance than the CO abundance, and therefore it will lead to a systematic increase in our values of $X_{\rm CO}$ compared to those for real clouds. The importance of this effect will depend on the size of the velocity dispersion in the cloud, but should not depend strongly on our choice of SFR, and it therefore represents a systematic error that biases all of our derived values of $X_{\rm CO}$ to slightly higher values. 

Another factor that may contribute to the discrepancy between our results and $X_{\rm CO, gal}$ is our choice to start the simulations with the hydrogen in fully molecular form.   
Although this approximates the behaviour that we find in previous studies of cloud formation \citep{clark12,smith14}, in practice the hydrogen in the assembling clouds is never {\em completely} molecular -- there is always some atomic component associated with the cloud. In equilibrium, we expect the surface density of atomic hydrogen associated with the clouds to be of order $10 \: {\rm M_{\odot} \: pc^{-2}}$ \citep{krum08,wolf10,stern14}, but the time taken to reach this equilibrium can be longer than the lifetime of the cloud. In our simulations with fully molecular initial conditions, we will therefore overestimate the H$_{2}$ content of the clouds by around 10-20\%, and hence will overestimate $X_{\rm CO}$ by a similar factor. In order to verify that the effect on $X_{\rm CO}$ is not greater than this, we ran two simulations of our high-mass cloud in which we started with all of the hydrogen in atomic form. In one of these simulations, we adopted the solar neighbourhood value for SFR (i.e.\ ${\rm SFR = SFR_{0}}$), while in the other we set ${\rm SFR = 100 \, SFR_{0}}$. The values of $X_{\rm CO}$ that we derived from these two runs are indicated in Figure~\ref{fig:xco_sfr} by the blue circles. We see that, as expected, the use of atomic rather than molecular initial conditions results in a decrease in $X_{\rm CO}$, but that the effect is not large and seems to be independent of our choice of SFR.

Finally, our decision to focus on clouds with a virial parameter $\alpha_{\rm vir} = 0.5$ may also lead to us deriving a larger value of $X_{\rm CO}$ than that found for large samples of Galactic GMCs. As discussed in the introduction, we know that the mean integrated CO intensity of a GMC depends sensitively on the CO line-width for that GMC \citep[see e.g.][]{Shetty2011a,Shetty2011b}. Increasing the line-width while keeping all other cloud properties the same leads to an increase in $W_{\rm CO}$ and hence a decrease in $X_{\rm CO}$. Similarly, decreasing the line-width decreases  $W_{\rm CO}$ and increases $X_{\rm CO}$. This implies that $X_{\rm CO}$ should depend to some extent on $\alpha_{\rm vir}$, in the sense that clouds with larger virial parameters that survive for long enough to form CO should have smaller values of $X_{\rm CO}$ than similar clouds with smaller values of $\alpha_{\rm vir}$. In any large sample of GMCs, we would not expect to find many clouds with $\alpha_{\rm vir} \ll 1$: any clouds that do initially have $E_{\rm kin} \ll E_{\rm grav}$ will undergo gravitational collapse, inducing motions that rapidly increase $E_{\rm kin}$ driving $\alpha_{\rm vir}$ towards 1 if the clouds continue to collapse or 0.5 if they settle into virial equilibrium \citep{bp11}. On the other hand, there is both theoretical \citep{dobbs11,wws14} and observational \citep{heyer09} evidence that many GMCs are gravitationally unbound, with $\alpha_{\rm vir} > 1$. It is therefore plausible that the mean value of $\alpha_{\rm vir}$ for the population of GMCs in the Milky Way is of order unity or larger. If so, then it follows that the clouds that we model in this study have systematically smaller CO line-widths than the average Galactic GMC, helping to explain why the values of $X_{\rm CO}$ that we derive for them are larger than $X_{\rm CO, gal}$.

\subsection{Dependence on the SFR}
\label{Sect:SFR}
We now come to the question at the heart of this paper: does the value of $X_{\rm CO}$ vary with the local ambient SFR? While the results displayed in Figure \ref{fig:xco_sfr} clearly show that $X_{\rm CO}$ is indeed a function of SFR, the extent to which the X-factor varies depends on several parameters, and we will now discuss these in turn.

\begin{figure}
\centerline{ \includegraphics[width=3.45in]{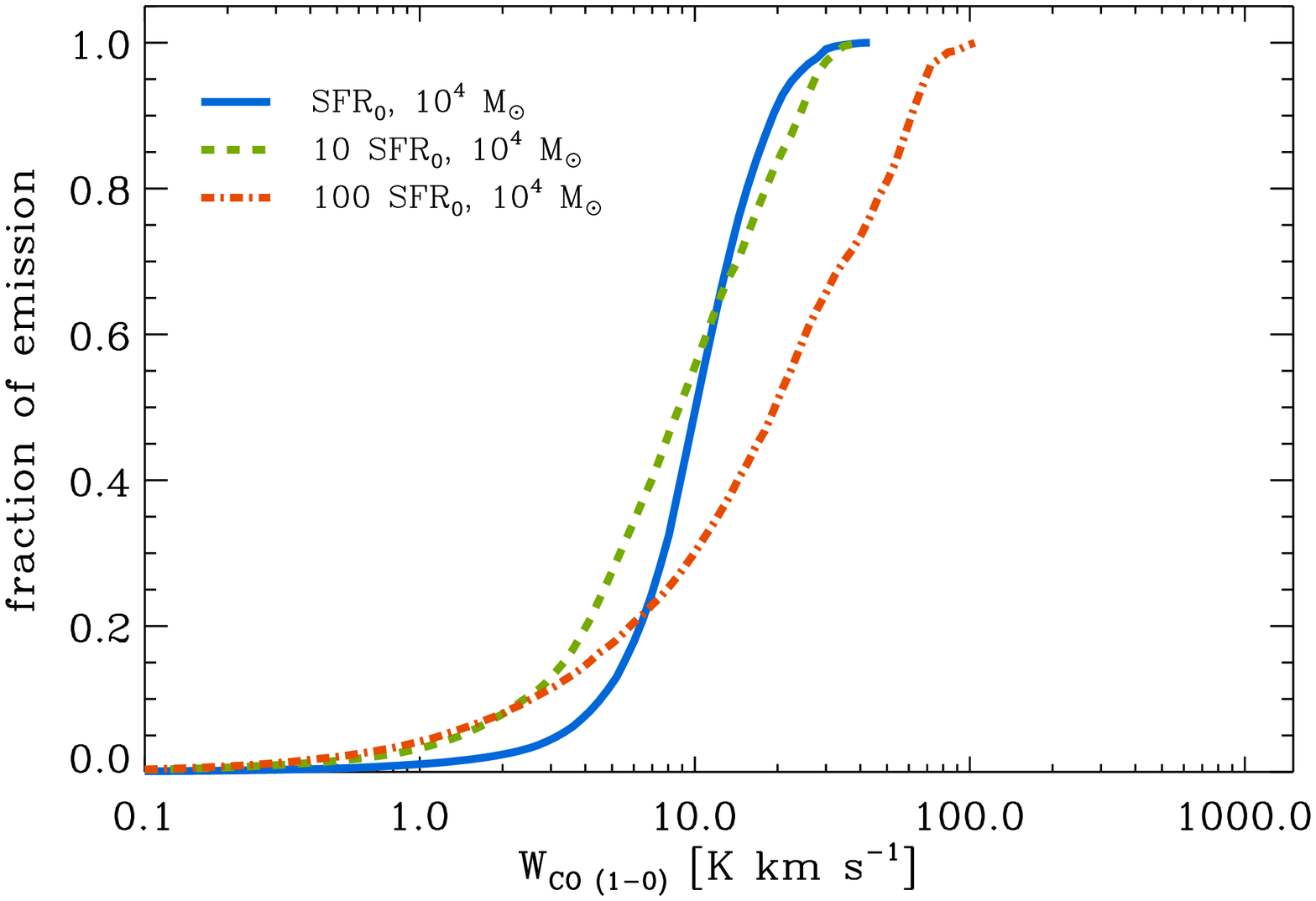}}
\centerline{ \includegraphics[width=3.45in]{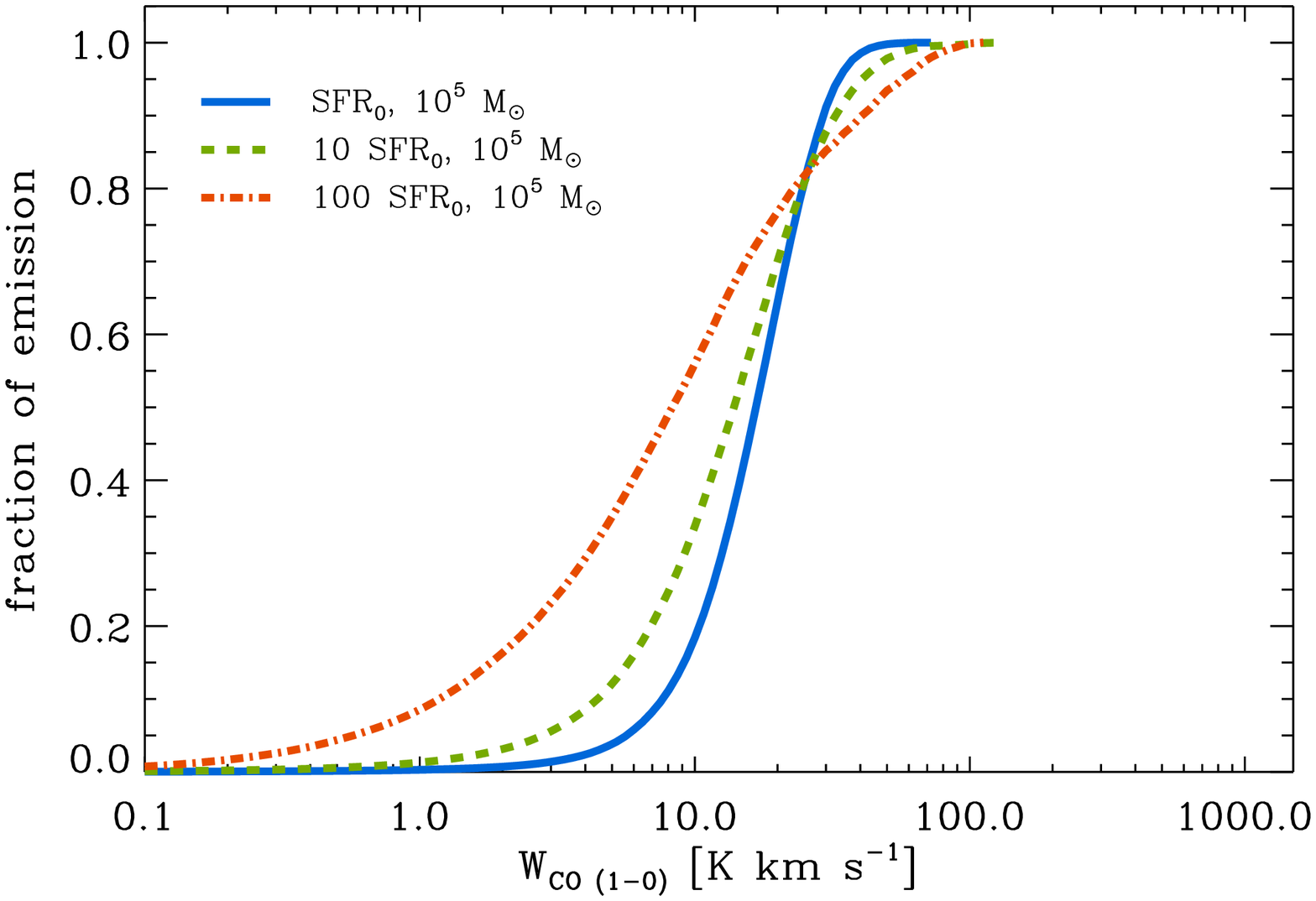}}
\centerline{ \includegraphics[width=3.45in]{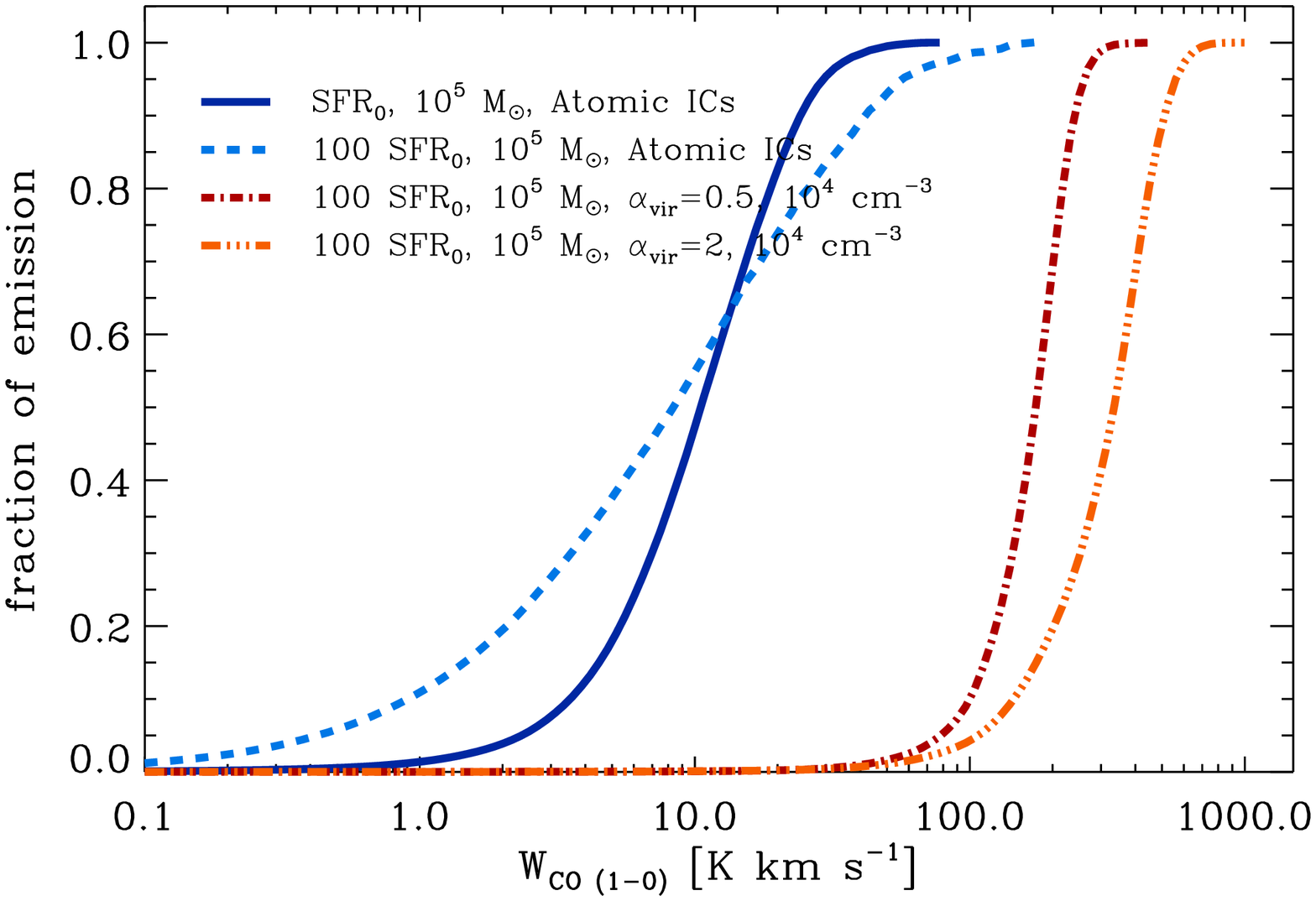}}
\caption{Fraction of the total emission that arises from pixels of a given $W_{\rm CO}$ and below.}
\label{fig:cumu_wco}
\end{figure}

First, examining the red circles in Figure \ref{fig:xco_sfr}, we see that, in general, the mean X-factor derived from the cloud maps increases as SFR increases. Given that CO is more easily dissociated than H$_2$, this makes sense: the CO is doing a progressively poorer job of tracing the true extent of the molecular gas as the ambient SFR increases. However we see that the effect is more pronounced for the high-mass clouds than for the low-mass clouds. Indeed, there is a slight decrease in the X-factor of the 100 SFR$_0$ case with respect to the 10 SFR$_0$ case for the low-mass clouds. So while the X-factor at solar neighbourhood values of the SFR (i.e.\ SFR$_0$) is independent of mass, this is not true when the SFR is increased.

The reason behind this behaviour in the low-mass clouds can be inferred from Figure~\ref{fig:cumu_wco}, where we show the fraction of the cloud's total CO emission that comes from pixels below a given $W_{\rm CO}$.  For progressively higher SFR, we see that fraction of the overall integrated emission coming from very bright lines of sight depend on the cloud mass. For the low-mass clouds at high SFR, the majority of the emission comes from regions with integrated intensities of 10 $\rm K\,km\,s^{-1}$ or greater, while the opposite is true for the high-mass clouds, in that they gain most of their emission from lines of sight with lower $W_{\rm CO}$. The idea summarised by \citet{bol13} that a few bright lines of sight can compensate for the lack of CO elsewhere in the cloud only appears to hold for our low-mass clouds.

The reason why the clouds behave differently has to do with their column density distributions. In the low-mass clouds, the mean column density is low enough that CO is efficiently photodissociated throughout much of their volume as the SFR is increased. The CO that survives is confined to the highly overdense regions produced by  turbulent compression and gravitational collapse, in which the shielding is more effective. Because of the high density, the excitation temperature of the CO in these regions is high, and so is the integrated intensity of the CO emission. These clouds therefore have many regions with very low $W_{\rm CO}$ and a few with very high $W_{\rm CO}$ that dominate the total emission. In the high-mass clouds, on the other hand, the mean column density of the gas is larger and the CO is therefore better able to resist photodissociation. Because of this, CO survives in lower density regions in these clouds even when the SFR is increased significantly. The excitation temperature of the CO in these regions is lower than in the dense cores, and so the integrated CO intensities coming from the lower density gas are smaller, but the much larger filling factor of these regions compensates for this, and leads to their contribution dominating the total emission from the cloud.

It is plausible that molecular clouds in environments with higher SFRs will be denser, on account of the higher ambient gas pressure in these environments \citep[see e.g.][]{elmegreen08,colom14}. Some support for this picture comes from observations of molecular clouds in the Central Molecular Zone of the Milky Way. The star formation rate in this region is roughly 50-100 times that in the solar neighbourhood \citep{bb11,longmore13a} and the molecular clouds located there have high mean densities, of the order of $10^4 \rm cm^{-3}$ or more \citep[see e.g.][]{dahmen98,longmore12,longmore13b}. There is also evidence to suggest that these clouds have higher values of $\alpha_{\rm vir}$ than local clouds \citep{kauf13,johnston14}.

To investigate this scenario, we performed two additional simulations that probe this regime. We modelled two clouds with mass $10^5 \rm M_{\odot}$ and initial number density $n_{0} = 10^4 \: {\rm cm^{-3}}$, in which the hydrogen was taken to be initially in molecular form. One of the clouds was initialized with a turbulent field resulting in a virial parameter $\alpha_{\rm vir} = 0.5$ (i.e.\ virial equilibrium), while the other had an $\alpha_{\rm vir} = 2$ (i.e.\ the cloud was gravitationally unbound). These are shown as the green circles in Figure \ref{fig:xco_sfr}. We see that the X-factors for these clouds are now much closer to the standard Milky Way value, demonstrating that the physical properties of the clouds play a significant role in controlling the X-factor. We also see that, as expected, the X-factor is lower in the case where the amount of turbulent kinetic energy in the cloud is higher. 

Another factor that strongly affects the mean X-factor derived from the cloud maps is the value used for deciding whether or not a pixel should be included in the averaging process, $W_{\rm CO, cut}$. If we set $W_{\rm CO, cut} = 0$, and hence include all of the lines of sight in our synthetic emission maps when determining  $\langle N_{\rm H_2} \rangle$ and $\langle W_{\rm CO} \rangle$, then we find that $X_{\rm CO}$ increases by between a factor of three (for the low-mass clouds) and an order of magnitude (for the high-mass clouds) as we increase the SFR from SFR$_{0}$ 
to 100~SFR$_{0}$. However, if we restrict our calculation of the X-factor to only those lines of sight with detectable CO emission, we find that the dependence of $X_{\rm CO}$ on the SFR becomes much weaker. For example, if we set $W_{\rm CO, cut} = 3\, \rm K\,km\,s^{-1}$, we find that the 
X-factor increases at most by a factor of two as we increase the SFR from SFR$_{0}$  to 100~SFR$_{0}$. This suggests that the main effect driving the increase in $X_{\rm CO}$ with SFR in the $W_{\rm CO, cut} = 0$ case is a large increase in the number of lines of sight that have molecular hydrogen but no detectable CO emission, or in other words that pass through ``CO-dark'' molecular gas \citep{wolf10}. In those regions of the clouds that remain CO-bright, the relationship between CO emission and H$_{2}$ column density varies only weakly with the SFR. This behaviour is similar to what is seen in observational studies of the dependence of $X_{\rm CO}$ on metallicity: observations that focus only on CO-bright regions find a weak dependence of $X_{\rm CO}$ on Z \citep[e.g.][]{wilson95,bolatto08}, while those that sample both CO-bright and CO-dark gas find a much stronger dependence 
\citep[e.g.][]{israel97,bolatto11}. Finally, it is also interesting to note that the X-factor for the high density, $100 \,\rm SFR_0$ clouds is insensitive to the value of $W_{\rm CO, cut}$. This is because these clouds are so dense that the ``skin'' in $A_{\rm V}$-space that contains H$_2$ but not CO is so thin that very little mass is contained within it. As such, the CO typically always does a good job of tracing the molecular content of these clouds.

\section{Discussion}
\subsection{Consequences for the Kennicutt-Schmidt relation}
The results presented in the previous section demonstrate that if we restrict our attention to those portions of molecular clouds 
that are traced by relatively bright CO emission, then the X-factor that we have to adopt in order to correctly convert from
the CO intensity of these regions to their H$_{2}$ column density does not vary strongly with the local environment.
Increasing the size of the assumed star formation rate (and hence the strength of the ISRF and the  cosmic ray ionisation rate)
by a factor of 100 leads to an increase in $X_{\rm CO}$ of around a factor of two at most. 
This result is consistent with the finding by \citet{pineda09} and \citet{hughes10} that in the LMC $X_{\rm CO}$ does not vary 
strongly with the strength of the ISRF, since these authors determine the masses of the clouds that they study using virial
mass estimates that are sensitive only to the properties of the CO-bright gas.

Unfortunately, the number of galaxies for which we can distinguish between CO-bright and CO-faint portions of GMCs is very
small. This is routinely done in the Milky Way, has already been done to some extent in the Magellanic Clouds, and with the
advent of ALMA becomes possible to do in more distant members of the Local Group.
However, in most extragalactic observations, the best spatial resolution that we can achieve
is comparable to or larger than the size of the individual clouds \citep[see e.g.][]{colom14}. In this case, we cannot easily
distinguish between CO-bright and CO-faint portions of the clouds, and so the results that we present for the case when
$W_{\rm CO, cut} = 0$ give the best guide to the behaviour of $X_{\rm CO}$ in these systems.

For these unresolved clouds, our results show that $X_{\rm CO}$ increases substantially as we increase the SFR, provided
that the other properties of the clouds remain unchanged (i.e.\ provided that their characteristic densities or velocity dispersions
do not also vary as functions of the SFR). One important implication of this is that in order to be able to use CO emission as a
reliable tracer of molecular mass, we need to understand the nature of the underlying cloud population and how this varies as
a function of the star formation rate. 

If we make the simplifying assumption that cloud properties do not depend to any great degree on the local value of the SFR,
then our results imply that the standard interpretation of the Kennicutt-Schmidt relation may need to be reassessed.
If the observed CO luminosity is dominated by high-mass clouds that have number densities similar to those of our fiducial 
cloud models, then our simulations suggest that $X_{\rm CO}$ should scale with the ambient rate of star formation as 
roughly\footnote{Note that this scaling is approximate: it is not a precise fit to the simulation results.}
\begin{equation}
\label{eq:ourxco}
X_{\rm CO} \propto \rm SFR^{1/2}.
\end{equation}
Provided that the SFR does not vary widely over the area of the observational beam, it then follows that
\begin{equation}
X_{\rm CO} \propto \rm \Sigma_{SFR}^{1/2}.  \label{xco_sfr}
\end{equation}
Measurements of the Kennicutt-Schmidt relation at gas surface densities where the molecular component dominates
\citep[e.g.][]{Bigiel2008,Blanc2009,Bigiel2011,liu11} find a power-law relationship that we can write as
\begin{equation}
\Sigma_{\rm SFR} \propto \Sigma_{\rm mol}^{N_{\rm obs}},
\end{equation}
where $N_{\rm obs}$ denotes the observed power-law index. In practice, however, $\Sigma_{\rm mol}$ is usually
not measured directly; instead, it is inferred from the CO luminosity through the application of some constant
CO-to-H$_{2}$ conversion factor, which in normal spiral galaxies is generally taken to have the canonical Galactic
value. Therefore, we can more accurately write the above relationship as
\begin{equation}
\Sigma_{\rm SFR} \propto \left(\Sigma_{\rm CO, em} X_{\rm CO, gal} \right)^{N_{\rm obs}},
\end{equation}
where $\Sigma_{\rm CO, em}$ is the surface density of CO emission. Now, if the actual CO-to-H$_{2}$ conversion factor
is not fixed, but instead varies with $\Sigma_{\rm SFR}$ according to Equation~\ref{xco_sfr}, then it follows that the
actual relationship between $\Sigma_{\rm SFR}$ and $\Sigma_{\rm mol}$ can be written as
\begin{equation}
\Sigma_{\rm SFR} \propto  \Sigma_{\rm mol}^{N_{\rm obs}} \Sigma_{\rm SFR}^{-N_{\rm obs}/2}.
\end{equation}
Rearranging this, we find that after correcting for the dependence of $X_{\rm CO}$ on the star formation
rate, the true relationship between $\Sigma_{\rm SFR}$ and $\Sigma_{\rm mol}$ can be written as
\begin{equation}
\Sigma_{\rm SFR} \propto \Sigma_{\rm mol}^{N_{\rm act}}
\end{equation}
where
\begin{equation}
N_{\rm act} = \frac{2 \, N_{\rm obs}}{2 + N_{\rm obs}}.
\end{equation}

Recent work by \citet{Bigiel2008} has suggested a value of $N_{\rm obs} = 1$ from a sample of nearby galaxies. In this case, the `true' index of the Kennicutt-Schmidt relation would be around $N_{\rm act} = 2/3$, implying that CO is a progressively worse tracer of the star formation rate as one moves to more extreme environments (see the discussion in \citealt{Shetty2014b}).  However, more recently, there have also been claims that $N_{\rm obs} < 1$ (i.e.\ that the Kennicutt-Schmidt relation is sub-linear), even when assuming a constant X-factor (e.g. \citealt{Ford2013, Shetty2013, Shetty2014a}), with the most likely value lying at around 0.76.  In this case, we would predict a value of $N_{\rm act} = 0.55$.  

It is important to note that this conclusion depends strongly on our assumption that the cloud properties do not vary strongly as a function of the local star formation rate. If this is not the case, and clouds in regions with higher values of $\Sigma_{\rm SFR}$ are systematically denser and/or more turbulent than clouds in less actively star-forming regions, then it is plausible that the decrease in $X_{\rm CO}$ caused by the higher densities and turbulent velocities could be large enough to overwhelm the increase caused by the higher star formation rate. Indeed, there is evidence that this is the case in the centres of many spiral galaxies: measurements of $X_{\rm CO}$ often find that it decreases close to the centre \citep[see e.g.][]{sandstrom13}, while we know from the study of the Central Molecular Zone of our own galaxy that molecular clouds in this environment are very dense and highly turbulent \citep[see e.g.\ the summary in the review of][]{molinari14}. Similarly, observational determinations of $X_{\rm CO}$ in nearby ultraluminous infra-red galaxies (ULIRGS) or rapidly star-forming systems at high redshift typically find values that are smaller than the canonical local value (see e.g.\ \citealt{ds98,magdis11,os11,hodge12,fu13} or the recent review by \citealt{casey14}). However, these observations probe regions with much higher gas surface densities and turbulent velocities than are found in the local ISM, once again suggesting that the decrease in $X_{\rm CO}$ caused by the increased densities and velocities is able to overcome the increase caused by the higher radiation field strength and cosmic ray ionization rate.

Ultimately, therefore, the real message to take away from this study is that it is necessary to understand the nature of the cloud population {\em before} one can properly interpret the CO emission. Unless one understands how the properties of the clouds depend on the local star formation rate, it is impossible to be certain whether $X_{\rm CO}$ increases with increasing $\Sigma_{\rm SFR}$ (as is the case in our simulations at fixed $\alpha_{\rm vir}$), decreases with increasing $\Sigma_{\rm SFR}$ (as appears to be the case in the centres of many spirals), or remains approximately constant. Further compounding this problem is the fact that CO measurements are one of the main observational tools used for determining the mass of molecular gas in clouds, potentially leaving us with a circular argument. Supplementary information on cloud masses, from tracers such as dust emission, can probably help to break this circularity, but a detailed examination of how best to do this is outside of the scope of this paper.

\subsection{Comparison with previous work}
It is interesting to compare our results on the dependence of the X-factor on the SFR with those of previous numerical studies that have looked at this problem. \citet{bell06} use the time-dependent {\sc ucl\_pdr} code \citep{papa02,bell05} to examine how $X_{\rm CO}$ varies as a function of depth within a semi-infinite slab of gas as they vary a number of different physical parameters, including the radiation field strength. They find that at high $A_{\rm V}$, increasing the radiation field strength has little effect on $X_{\rm CO}$: changing $G_{0}$ by a factor of $10^{6}$ leads to no more than a factor of three change in $X_{\rm CO}$. At low $A_{\rm V}$, on the other hand, the dependence of $X_{\rm CO}$ on the radiation field is much stronger. At very low $A_{\rm V}$, the gas is CO-faint and $X_{\rm CO}$ is very large, but as we move to higher $A_{\rm V}$, the CO content of the gas grows, and the value of $X_{\rm CO}$ approaches the canonical Galactic value. The depth into the cloud at which this transition occurs depends on the radiation field strength, with stronger fields implying that the transition occurs at higher $A_{\rm V}$. It is not straightforward to convert from these results to a cloud-averaged $X_{\rm CO}$, but it is clear that they are at least qualitatively consistent with our findings that $X_{\rm CO}$ measured only for CO-bright regions does not vary strongly with the SFR,  while $X_{\rm CO}$ averaged over the whole cloud (including the CO-faint regions) shows a much stronger dependence.

\citet{feld12} have also explored the dependence of $X_{\rm CO}$ on the local radiation field. They performed a large hydrodynamical simulation of the formation of a typical $L_{*}$ galaxy \citep[see also][]{gk11} and then post-processed the results of this simulation using a sub-grid prescription for $X_{\rm CO}$ that assumes that it is primarily determined by the mean visual extinction of the individual clouds and the strength of the ISRF. For $G_{0} = 1$, \citet{feld12} use the results of \citet{gm11} to calibrate their subgrid model, while for $G_{0} \gg 1$ and $G_{0} \ll 1$, they estimate the CO abundance (and hence $X_{\rm CO}$) using a simple remapping procedure that assumes that the clouds are in photodissociation equilibrium. 
They find that for clouds with mean extinctions $\bar{A}_{\rm V} \sim 6$ and below, the value of $X_{\rm CO}$ depends strongly on the radiation field strength. For higher column density clouds, on the other hand, $X_{\rm CO}$ becomes largely independent of the radiation field strength. As most of the molecular clouds that form in their simulations have relatively large column densities, they find that when you average over the whole population of clouds, the dependence of $X_{\rm CO}$ on the strength of the ISRF becomes very weak.

At first sight, these results would appear to contradict our finding that $X_{\rm CO}$ can vary significantly as we change the SFR. However, there are a couple of important points that one should bear in mind. First, \citet{feld12} do not vary the cosmic ray ionisation rate, only the radiation field strength. Therefore, their models do not account for the increased destruction of CO by dissociative charge transfer with He$^{+}$ that occurs when the cosmic ray ionisation rate is large. At high $A_{\rm V}$, this is the main CO destruction mechanism, and its inclusion in our models is one of the main reasons why we see at least some dependence of $X_{\rm CO}$ on SFR even when we restrict our attention to portions of the cloud that are highly shielded and CO-bright. Second, the hydrodynamical simulations used as a basis for the \citet{feld12} analysis have a resolution of only 60~pc and hence are sensitive only to the largest, densest clouds, which are naturally the ones least affected by changes in the SFR.

Another large-scale numerical study of the dependence of $X_{\rm CO}$ on environment was presented by \citet{desika12}. They post-processed an extensive series of SPH simulations of isolated and merging galaxies in order to determine the local chemical and thermal state of the gas and the consequent CO emission (see also \citealt{klm11} and \citealt{desika11}).  They found that in their models, $X_{\rm CO}$ scales inversely with the molecular gas surface density as $X_{\rm CO} \propto \Sigma_{\rm H_{2}}^{-1/2}$. In these models, $\Sigma_{\rm SFR} \propto \Sigma_{\rm H_{2}}^{3/2}$ by construction, and so the implication is that $X_{\rm CO}$ should decrease weakly with increasing star formation rate, scaling as $X_{\rm CO} \propto \Sigma_{\rm SFR}^{-1/3}$. However, as with many of the observational studies of $X_{\rm CO}$ in extreme environments mentioned in the previous section, direct comparison of these results with our own is complicated by the fact that the mean properties of the clouds in regions with high SFR in these simulations are not the same as those of the clouds in regions with low SFR. 

\citet{lagos12} also studied the effects of changes in the strength of the ISRF and the cosmic ray ionization rate on the value of $X_{\rm CO}$ using a simple semi-analytical approach. They used the {\sc galform} semi-analytical galaxy formation model \citep{lagos11a,lagos11b} to compute the atomic and molecular gas content of a series of model galaxies, and then used the {\sc ucl\_pdr} code to determine the CO luminosity of the molecular gas, from which they could then infer $X_{\rm CO}$. In their PDR calculations, it was necessary for them to assume some representative number density and visual extinction for the molecular gas. In most of their models, \citet{lagos12} assumed a characteristic number density $n = 10^{4} \: {\rm cm^{-3}}$ and a characteristic extinction $A_{\rm V} = 8$. In these conditions, one would expect CO photodissociation to be completely negligible even when the ISRF is very strong, meaning that changes in the radiation field strength will only affect $X_{\rm CO}$ indirectly, through its effect on the thermal balance of the dust. Indeed, \citet{lagos12} find that in their model, even very large changes in the radiation field strength have only a small influence on $X_{\rm CO}$. However, 
in practice, in realistic GMC models only a small fraction of the total gas mass is found in regions with a mean visual extinction as high as $A_{\rm V} = 8$ \citep[see e.g.][]{cg14}, and so by focussing only on these conditions, \citet{lagos12} ignore the large changes in the CO content and CO luminosity of the lower density, less shielded gas that drive much of the variation that we find in $X_{\rm CO}$.  \citet{lagos12} also examine the effect of varying the cosmic ray ionization rate while keeping the radiation field strength fixed, and find that in this case, $X_{\rm CO}$ increases slowly with increasing ionization rate, in general agreement with our results.

The dependence of $X_{\rm CO}$ on the UV field strength was also examined by \citet{offner14}. They computed $X_{\rm CO}$ for clouds illuminated by radiation fields with 1 and 10 times the strength of the \citet{dr78} field, using the {\sc 3d-pdr} code \citep{bisbas12}. When computing $X_{\rm CO}$, they only considered pixels in their synthetic emission maps with CO integrated intensities $W_{\rm CO} > 0.45 \: {\rm K \: km \: s^{-1}}$. They find that $X_{\rm CO} = 1.5 X_{\rm CO, gal}$ for their $G_{0} = 1.7$ run and $X_{\rm CO} = 2.15 X_{\rm CO, gal}$ for their $G_{0} = 17$ run, a roughly 50\% increase. This is somewhat smaller than the difference we find between our ${\rm SFR = SFR_{0}}$ and ${\rm SFR = 10 \, SFR_{0}}$ runs, which for a similar integrated intensity cut show values of $X_{\rm CO}$ that differ by closer to a factor of two. However, it is important to note that \citet{offner14} do not vary the cosmic ray ionisation rate at the same time that they vary the radiation field strength, and so it is likely that there is less destruction of CO in their $G_{0} = 17$ run than in our ${\rm SFR = 10 \, SFR_{0}}$ run.

Finally, in a recent paper, \citet{bpv15} have studied how the typical CO abundance in a GMC varies as a function of the cosmic ray ionisation rate. They find that increasing the cosmic ray ionisation rate by a factor of ten or more leads to substantial destruction of CO and argue that this will lead to clouds in regions with ${\rm SFR > 10 \, SFR_{0}}$ being CO-faint. However, they do not compute $X_{\rm CO}$ for any of their clouds models and hence do not quantify its dependence on the SFR.

\subsection{Caveats}
There are two important methodological caveats that the reader should bear in mind when considering our results. First, as we have already discussed in Section~\ref{sfr0}, we do not currently account for the effects of line-of-sight velocity gradients when determining the effectiveness of H$_{2}$ self-shielding. This means that we will tend to overestimate the effectiveness of self-shielding in gas with H$_{2}$ column densities of around $10^{14} < N_{\rm H_{2}} < 10^{18} \: {\rm cm^{-2}}$, the regime where Doppler broadening of the Lyman-Werner lines dominates the UV absorption spectrum. Consequently, we will overestimate the total H$_{2}$ content of the cloud. However, as most of the H$_{2}$ in the cloud is found in regions with much higher H$_{2}$ column densities, we would not expect the omission of this effect to have a large impact on our results.

Second, we have assumed that the cosmic ray ionisation rate is uniform throughout the cloud, or in other words that the cosmic ray absorption spectrum is not significantly affected by absorption within the cloud. Whether or not this is a good approximation depends on the details of the low energy portion of the cosmic ray energy spectrum, which is poorly constrained 
\citep{pgg09}. There is some tentative evidence from astrochemical studies that the cosmic ray ionisation rate in dense cores may be significantly lower than in the diffuse ISM \citep[see e.g.][]{ind12}, but the issue is far from being settled. If there is indeed a significant fall-off in the cosmic ray ionisation rate as we move from diffuse to dense gas, then our models will tend to overestimate the heating rate in the dense gas, and also the rate at which CO is destroyed there. The net effect of this on $X_{\rm CO}$ is difficult to predict without detailed modelling. However, we note that the destruction of the relatively diffuse inter-clump CO component that occurs as we increase the SFR is a consequence of the increasing strength of the ISRF, and not the increase in the cosmic ray ionisation rate. Therefore, even if cosmic ray absorption within the cloud becomes significant, this behaviour will remain the same, and so we would expect to still see an increase in $X_{\rm CO}$ as we increase the SFR.

In addition to these methodological caveats, we also take this opportunity to remind the reader that when we refer to changes in the ``SFR'', what we mean are changes in the strength of the ISRF and the size of the cosmic ray ionisation rate, since we assume that both scale linearly with the star formation rate. It could well be that this assumption is too simplistic, and that the functional dependence of the radiation field strength and the cosmic ray ionisation rate on the local SFR is more complicated than a simple linear scaling. If so, then the results of our study will still hold, but the actual star formation rates corresponding to our $10 \, {\rm SFR_{0}}$ and $100 \, {\rm SFR_{0}}$ models will differ from what we have assumed here.

\section{Conclusions}
In this paper, we have performed a series of numerical simulations of molecular cloud evolution that explore how the distribution of CO emission coming from the clouds changes as we change the strength of the interstellar radiation field and the size of the cosmic ray ionisation rate in the clouds. If one makes the reasonable assumption that these quantities scale linearly with the local star formation rate, then our results indicate how the CO-to-H$_{2}$ conversion factor for the clouds, $X_{\rm CO}$, depends on the star formation rate.

We find that as we increase the radiation field strength and the cosmic ray ionisation rate, the CO content of our simulated clouds decreases. CO survives more effectively in well-shielded clump and filaments than in the more diffuse inter-clump gas, leading to a significant decrease in the filling factor of bright CO emission. The integrated intensity of the CO emission from the brightest regions increases with increasing SFR, owing to the heating of the gas by the higher cosmic ray flux, but this increase is too small to compensate for the loss of CO emission from elsewhere in the cloud, and so overall the CO luminosity of the cloud decreases. 

The change in the chemical composition of the cloud as we increase the SFR leads to a change in $X_{\rm CO}$. However, the size of this change depends on the method we use to compute $X_{\rm CO}$. If we consider the whole area of the cloud when computing the mean H$_{2}$ column density and CO integrated intensity, then we find that $X_{\rm CO}$ changes substantially, scaling as roughly $X_{\rm CO} \propto {\rm SFR}^{1/2}$ for our high-mass clouds. On the other hand, if we consider only regions with CO integrated intensities 
exceeding an intensity threshold $W_{\rm CO, cut}$, then we find that the dependence of $X_{\rm CO}$ on SFR weakens substantially as we increase $W_{\rm CO, cut}$. 
The reason for this difference in behaviour is that the fraction of the cloud filled with ``dark'' molecular gas (i.e.\ H$_{2}$ without associated CO) increases as we increase the SFR. This dark gas contributes to our calculation of $X_{\rm CO}$ in the case where we do not apply an intensity threshold, but is ignored when we do apply a threshold. 

We have also explored whether increasing the density and turbulent velocity dispersion of our model clouds at the same time as we increase the SFR affects the relationship between $X_{\rm CO}$ and the SFR. We find that if we increase the initial number density to $n_{0} = 10^{4} \: {\rm cm^{-3}}$ and raise the virial parameter to $\alpha_{\rm vir} = 2$, then we can recover a value of $X_{\rm CO}$ close to the canonical Galactic value even when the local SFR is 100 times our default value. We conclude from this that we will only recover a positive correlation between $X_{\rm CO}$ and the SFR if there is not a strong correlation between mean cloud densities and the SFR. The fact that observational determinations of $X_{\rm CO}$ in starbursting systems such as ULIRGs find a value {\em lower} than the canonical Galactic value therefore provides a strong indication that the properties of typical molecular clouds in these systems must differ significantly from those of molecular clouds in the local ISM. Consequently, we argue that the most important message to take away from this study is that it is necessary to understand the nature of the cloud population before one can properly interpret the CO emission coming from the clouds. 

Finally, we stress that if the CO emission from nearby galaxies is dominated by virialised clouds with masses around $10^5$ $\rm M_{\odot}$ and densities around a few 100 ${\rm cm}^{-3}$, then the X-factor may vary significantly as the local SFR changes. In these circumstances, our standard interpretation of the Kennicutt-Schmidt relation, in which $X_{\rm CO}$ is assumed to be constant throughout the galaxy, should be rethought. 

\section*{Acknowledgements}
The authors would like to thank J. Pety, R.~Feldmann, N.~Gnedin, and F.~Israel for interesting discussions concerning the physics of the CO-to-H$_{2}$ conversion factor in strong radiation fields. They would also like to thank the referee, D.~Narayanan, for a very helpful and constructive report that helped to improve the paper. SCOG acknowledges financial support from the Deutsche Forschungsgemeinschaft  via SFB 881, ``The Milky Way System'' (sub-projects B1, B2 and B8) and SPP 1573, ``Physics of the Interstellar Medium'' (grant number GL 668/2-1).


\begin{thebibliography}{}

\bibitem[Ballesteros-Paredes \& {Mac Low}(2002)]{bpm02}
Ballesteros-Paredes, J., \& {Mac Low}, M.-M.\ 2002, ApJ, 570, 734

\bibitem[Ballesteros-Paredes et~al.(2011)]{bp11}
Ballesteros-Paredes, J., Hartmann, L.~W., V\'azquez-Semadeni, E., Heitsch, F., \& Zamora-Avil\'es, M.~A.\ 2011,
MNRAS, 411, 65

\bibitem[Bate \& Burkert(1997)]{bb97}
Bate, M.~R., \& Burkert, A. 1997, MNRAS, 288, 1060

\bibitem[Beaumont et~al.(2013)]{beau13}
Beaumont, C.~N., Offner, S.~S.~R., Shetty, R., Glover, S.~C.~O., \& Goodman, A.~A.\ 2013, ApJ, 777, 173

\bibitem[Bell et~al.(2005)]{bell05}
Bell, T.~A., Viti, S., Williams, D.~A., Crawford, I.~A., \& Price, R.~J.\ 2005, MNRAS, 357, 961

\bibitem[Bell et~al.(2006)]{bell06}
Bell, T.~A., Roueff, E., Viti, S., \& Williams, D.~A.\ 2006, MNRAS, 371, 1865

\bibitem[Bergin \& Tafalla(2007)]{bt07}
Bergin, E.~A., \& Tafalla, M. 2007, ARA\&A, 45, 339

\bibitem[Bigiel et al.(2008)]{Bigiel2008} Bigiel, F., Leroy, A., 
Walter, F., et al.\ 2008, AJ, 136, 2846

\bibitem[Bigiel et~al.(2011)]{Bigiel2011}
Bigiel, F., et~al.\ 2011, ApJ, 730, L13

\bibitem[Bisbas et~al.(2012)]{bisbas12}
Bisbas, T.~G., Bell, T.~A., Viti, S., Yates, J., \& Barlow, M.~J.\ 2012, MNRAS, 427, 2100 

\bibitem[Bisbas, Papadopoulos \& Viti(2015)]{bpv15}
Bisbas, T.~G., Papadopoulos, P.~P., \& Viti, S.\ 2015, ApJ, 803, 37

\bibitem[Black(1994)]{bl94}
Black, J.~H. 1994, ASP Conf.\ Ser.\ 58, in The First Symposium on the Infrared Cirrus
and Diffuse Interstellar Clouds, eds.\ R.~M.~Cutri \& W.~B.~Latter, (San Francisco:ASP), 355

\bibitem[Blanc et~al.(2009)]{Blanc2009}
Blanc, G.~A., Heiderman, A., Gebhardt, K., Evans, N.~J., \& Adams, J.\ 2009, ApJ, 704, 842 

\bibitem[Blitz et~al.(2007)]{blitz07}
Blitz, L., Fukui, Y., Kawamura, A., Leroy, A., Mizuno, N., \& Rosolowsky, E.\ 2007, in Protostars \& Planets V,
eds.\ B.~Reipurth, D.~Jewitt, \& K.~Keil, (Tucson: University of Arizona Press), 81

\bibitem[Bolatto et~al.(2008)]{bolatto08}
Bolatto, A.~D., Leroy, A.~K., Rosolowsky, E., Walter, F., \& Blitz, L.\ 2008, ApJ, 686, 948

\bibitem[Bolatto et~al.(2011)]{bolatto11}
Bolatto, A.~D., et~al., 2011, ApJ, 741, 12

\bibitem[Bolatto, Wolfire \& Leroy(2013)]{bol13}
Bolatto, A.~D., Wolfire, M.,  \& Leroy, A.~K.\ 2013, ARA\&A, 51, 207

\bibitem[Bonatto \& Bica(2011)]{bb11}
Bonatto, C., \& Bica, E.\ 2011, MNRAS, 415, 2827

\bibitem[Casey, Narayanan \& Cooray(2014)]{casey14}
Casey, C.~M., Narayanan, D., \& Cooray, A.\ 2014, Phys.\ Rep., 541, 45

\bibitem[Clark, Glover \& Klessen(2012)]{cgk12}
Clark, P.~C., Glover, S.~C.~O., \& Klessen, R.~S.\ 2012, MNRAS, 420, 745

\bibitem[Clark et al.(2012)]{clark12}
Clark, P.~C., Glover, S.~C.~O., Klessen, R.~S., \& Bonnell, I.~A.\ 2012, MNRAS, 424, 2599

\bibitem[Clark \& Glover(2014)]{cg13}
Clark, P.~C., \& Glover, S.~C.~O.\ 2014, MNRAS, 444, 2396

\bibitem[Clark et~al.(2013)]{clark13}
Clark, P.~C., Glover, S.~C.~O., Ragan, S.~E., Shetty, R., \& Klessen, R.~S.\ 2013, ApJ, 768, L34

\bibitem[Clark \& Glover(2014)]{cg14}
Clark, P.~C., \& Glover, S.~C.~O., 2014, MNRAS, 444, 2396

\bibitem[Colombo et~al.(2014)]{colom14}
Colombo, D., et~al., 2014, ApJ, 784, 3

\bibitem[Dahmen et~al.(1998)]{dahmen98}
Dahmen, G., Huttemeister, S., Wilson, T.~L., Mauersberger, R.\ 1998, A\&A, 331, 959

\bibitem[Dame et~al.(2001)]{dame01}
Dame, T.~M., Hartmann, D., \& Thaddeus, P. 2001, ApJ, 547, 792

\bibitem[Dickman(1978)]{dick78}
Dickman, R.~L. 1978, ApJS, 37, 407

\bibitem[Digel et~al.(1999)]{digel99}
Digel, S.~W., Aprile, E., Hunter, S.~D., Mukherjee, R., \& Xu, F. 1999, ApJ, 520, 196

\bibitem[Dobbs, Burkert \& Pringle(2011)]{dobbs11}
Dobbs, C.~L., Burkert, A., \& Pringle, J.~E.\ 2011, MNRAS, 413, 2935

\bibitem[Downes \& Solomon(1998)]{ds98}
Downes, D., \& Solomon, P.~M.\ 1998, ApJ, 507, 615

\bibitem[Draine(1978)]{dr78}
Draine, B.~T. 1978, ApJS,  36, 595

\bibitem[Elmegreen, Klessen \& Wilson(2008)]{elmegreen08}
Elmegree, B.~G., Klessen, R.~S., \& Wilson, C.~D.\ 2008, ApJ, 681, 365

\bibitem[Feldmann et~al.(2012)]{feld12} 
Feldmann, R., Gnedin, N.~Y., \& Kravtsov, A.~V., 2012, ApJ, 747, 124

\bibitem[Ferriere(2001)]{fer01}
Ferriere, K.~M.\ 2001, Rev.\ Mod.\ Phys., 73, 1031

\bibitem[Ford et al.(2013)]{Ford2013} Ford, G.~P., Gear, W.~K., 
Smith, M.~W.~L., et al.\ 2013, ApJ, 769, 55

\bibitem[Froebrich \& Rowles(2010)]{fr10}
Froebrich, D., \& Rowles, J. 2010, MNRAS, 406, 1350

\bibitem[Fu et~al.(2013)]{fu13}
Fu, H., et~al., 2013, Nature, 498, 338

\bibitem[Glover \& {Mac Low}(2007a)]{gm07a}
Glover, S.~C.~O., \& {Mac Low}, M.-M.\ 2007a, ApJS, 169, 239

\bibitem[Glover \& {Mac Low}(2007b)]{gm07b}
Glover, S.~C.~O., \& {Mac Low}, M.-M.\ 2007b, ApJ, 659, 1317

\bibitem[Glover et al.(2010)]{glo10}
Glover, S.~C.~O., Federrath, C., {Mac Low}, M.-M., \& Klessen, R.~S.\ 2010, MNRAS, 404, 2

\bibitem[Glover \& Clark(2012a)]{gc12a}
Glover, S.~C.~O., \& Clark, P.~C.\ 2012a, MNRAS, 421, 116

\bibitem[Glover \& Clark(2012b)]{gc12b}
Glover, S.~C.~O., \& Clark, P.~C.\ 2012b, MNRAS, 426, 377


\bibitem[Glover \& {Mac Low}(2011)]{gm11}
Glover, S.~C.~O., \& {Mac Low}, M.-M. 2011, MNRAS, 412, 337

\bibitem[Gnedin \& Kravtsov(2011)]{gk11}
Gnedin, N.~Y., \& Kravtsov, A.~V.\ 2011, ApJ, 728, 88

\bibitem[Goldsmith(2001)]{gold01}
Goldsmith, P.~F. 2001, ApJ, 557, 736

\bibitem[Goodman et~al.(2009)]{good09}
Goodman, A.~A., Pineda, J.~E., \& Schnee, S.~L.\ 2009, ApJ, 692, 91

\bibitem[Gredel, Lepp \& Dalgarno(1987)]{gld87}
Gredel, R., Lepp, S., \& Dalgarno, A.\ 1987, ApJ, 323, L137

\bibitem[Habing(1968)]{habing68}
Habing, H.~J. 1968, Bull. Astron. Inst. Netherlands,  19, 421

\bibitem[Heyer et~al.(2009)]{heyer09}
Heyer, M., Krawczyk, C., Duval, J., \& Jackson, J.~M.\ 2009, ApJ, 699, 1092

\bibitem[Hodge et~al.(2012)]{hodge12}
Hodge, J.~A., Carilli, C.~L., Walter, F., {de Blok}, W.~J.~G., Riechers, D., Daddi, E., \& Lentati, L.\
2012, ApJ, 760, 11

\bibitem[Hughes et~al.(2010)]{hughes10}
Hughes, A., et~al.\ 2010, MNRAS, 406, 2065

\bibitem[Indriolo \& McCall(2012)]{ind12}
Indriolo, N., \& McCall, B.~J.\ 2012, ApJ, 745, 91

\bibitem[Israel(1997)]{israel97}
Israel, F.~P.\ 1997, A\&A, 328, 471

\bibitem[Johnston et~al.(2014)]{johnston14}
Johnston, K.~G., Beuther, H., Linz, H., Schmiedeke, A., Ragan, S.~E., \& Henning, Th.\ 2014, A\&A, 568, A56

\bibitem[Kainulainen et~al.(2009)]{kain09}
Kainulainen, J., Beuther, H., Henning, T., Plume, R. 2009, A\&A, 508, L35

\bibitem[Kauffmann, Pillai \& Zhang(2013)]{kauf13}
Kauffmann, J., Pillai, T., \& Zhang, Q.\ 2013, ApJ, 765, L35

\bibitem[Krumholz et~al.(2008)]{krum08}
Krumholz, M.~R., McKee, C.~F., \& Tumlinson, J.\ 2008, ApJ, 689, 865

\bibitem[Krumholz, Leroy \& McKee(2011)]{klm11}
Krumholz, M.~R., Leroy, A.~K., \& McKee, C.~F.\ 2011, ApJ, 731, 25

\bibitem[Lagos et~al.(2011a)]{lagos11a}
Lagos, C.~D., Lacey, C.~G., Baugh, C.~M., Bower, R.~G., \& Benson, A.~J.\ 2011a, MNRAS, 416, 1566

\bibitem[Lagos et~al.(2011b)]{lagos11b}
Lagos, C.~D., Baugh, C.~M., Lacey, C.~G., Benson, A.~J., Kim, H.-S., Power, C.\ 2011b, MNRAS, 418, 1649

\bibitem[Lagos et~al.(2012)]{lagos12}
Lagos, C.~D., et~al., 2012, MNRAS, 426, 2142

\bibitem[Larson(1981)]{larson81}
Larson, R.~B., 1981, MNRAS, 194, 809

\bibitem[Lee et~al.(1996)]{lee96}
Lee, H.-H., Herbst, E., {Pineau des For\^ets}, G.,
Roueff, E., \& {Le Bourlot}, J.\ 1996, A\&A, 311, 690

\bibitem[Lee et~al.(2014)]{myl14}
Lee, M.-Y., Stanimirovi\'c, S., Wolfire, M.~G., Shetty, R., Glover, S.~C.~O.,
Molina, F.~Z., \& Klessen, R.~S.\ 2014, ApJ, 784, 80

\bibitem[Liu et~al.(2011)]{liu11}
Liu, G., Koda, J., Calzetti, D., Fukuhara, M., \& Momose, R.\ 2011, ApJ, 735, 63

\bibitem[Longmore et~al.(2012)]{longmore12}
Longmore, S., et~al., 2012, ApJ, 746, 117

\bibitem[Longmore et~al.(2013a)]{longmore13a}
Longmore, S., et~al., 2013a, MNRAS, 429, 987

\bibitem[Longmore et~al.(2013b)]{longmore13b}
Longmore, S., et~al., 2013b, MNRAS, 433, L15

\bibitem[Magdis et~al.(2011)]{magdis11}
Magdis, G.~E.,  et~al., 2011, ApJ, 740, L15

\bibitem[Molina, Glover \& Federrath(2011)]{molina11}
Molina, F., Glover, S.~C.~O., \& Federrath, C.\ 2011, in Conditions and Impact of Star Formation, 
EAS Publications Series Volume 52, eds.\ M.~R\"ollig, R.~Simon, V.~Ossenkopf  \& J.~Stutzki, p.\ 289

\bibitem[Molinari et~al.(2014)]{molinari14}
Molinari, S., Bally, J., Glover, S., Moore, T., Noreiga-Crespo, A., Plume, R., Testi, L., V\'azquez-Semadeni, E.,
Zavagno, A., Bernard, J.-P., \& Martin P.\ 2014, in Protostars and Planets VI, eds.\ H.~Beuther, R.~S.~Klessen,
C.~P.~Dullemond and T.~Henning, (Tucson: University of Arizona Press), 125

\bibitem[Narayanan et~al.(2011)]{desika11}
Narayanan, D., Krumholz, M.~R., Ostriker, E.~C., \& Hernquist, L.\ 2011, MNRAS, 418, 664

\bibitem[Narayanan et~al.(2012)]{desika12}
Narayanan, D., Krumholz, M.~R., Ostriker, E.~C., \& Hernquist, L.\ 2012, MNRAS, 421, 3127

\bibitem[Narayanan \& Hopkins(2013)]{nh13}
Narayanan, D., \& Hopkins, P.~F.\ 2013, MNRAS, 433, 1223

\bibitem[Nelson \& Langer(1999)]{nl99}
Nelson, R.~P., \& Langer, W.~D. 1999, ApJ, 524, 923

\bibitem[Offner et~al.(2014)]{offner14}
Offner, S.~S.~R., Bisbas, T.~G., Bell, T.~A., \& Viti, S.\ 2014, MNRAS, 440,  L81

\bibitem[Ostriker \& Shetty(2011)]{os11}
Ostriker, E.~C., \& Shetty, R.\ 2011, ApJ, 731, 41

\bibitem[Padovani, Galli \& Glassgold(2009)]{pgg09}
Padovani, M., Galli, D., \& Glassgold, A.~E.\ 2009, A\&A, 501, 619

\bibitem[Papadopoulos et~al.(2002)]{papa02}
Papadopoulos, P.~P., Thi, W.-F., \& Viti, S.\ 2002, ApJ, 579, 270

\bibitem[Pineda, Caselli \& Goodman(2008)]{pineda08}
Pineda, J.~E., Caselli, P., \& Goodman, A.~A.\ 2008, ApJ, 679, 481

\bibitem[Pineda et~al.(2009)]{pineda09}
Pineda, J.~L., Ott, J., Klein, U., Wong, T., Muller, E., \& Hughes, A.\ 2010, ApJ, 703, 736

\bibitem[Pineda et~al.(2010)]{pineda10}
Pineda, J.~L., Goldsmith, P.~F., Chapman, N., Snell, R.~L., Li, D., Cambresy, L. \& Brunt, C. 2010,
ApJ, 721, 686

\bibitem[Roman-Duval et~al.(2010)]{rd10}
Roman-Duval, J., Jackson, J.~M., Heyer, M., Rathborne, J., \& Simon, R.,
2010, ApJ, 723, 492

\bibitem[Sanders et~al.(1984)]{sand84}
Sanders, D.~B., Solomon, P.~M., Scoville, N.~Z. 1984,
ApJ, 276, 182

\bibitem[Sandstrom et~al.(2013)]{sandstrom13}
Sandstrom, K.~M., et~al.\ 2013, ApJ, 777, 5

\bibitem[Sch\"oier et~al.(2005)]{sch05}
Sch\"oier, F.~L., {van der Tak}, F.~F.~S., {van Dishoeck}, E.~F., \& Black, J.~H.\ 2005, A\&A, 432, 369

\bibitem[Scoville et~al.(1987)]{sco87}
Scoville, N.~Z., Yun, M.~S., Sanders, D.~B., Clemens, D.~P., \&
Waller, W.~H.\ 1987, ApJS, 63, 821

\bibitem[Sembach et~al.(2000)]{sembach00}
Sembach, K.~R., Howk, J.~C., Ryans, R.~S.~I., \& Keenan, F.~P.\ 2000, ApJ, 528, 310

\bibitem[Shetty et al.(2011a)]{Shetty2011a} 
Shetty, R., Glover, S.~C., Dullemond, C.~P., \& Klessen, R.~S.\ 2011a, MNRAS, 412, 1686

\bibitem[Shetty et al.(2011b)]{Shetty2011b} 
Shetty, R., Glover, S.~C., Dullemond, C.~P.,  Ostriker, E.~C., Harris, A.I., Klessen, R.~S.\ 2011b, MNRAS, 415, 3253

\bibitem[Shetty et al.(2013)]{Shetty2013} Shetty, R., Kelly, 
B.~C., \& Bigiel, F.\ 2013, MNRAS, 430, 288

\bibitem[Shetty et al.(2014a)]{Shetty2014a} Shetty, R., Kelly, 
B.~C., Rahman, N., et al.\ 2014a, MNRAS, 437, L61

\bibitem[Shetty et al.(2014b)]{Shetty2014b} Shetty, R., Clark, 
P.~C., \& Klessen, R.~S.\ 2014b, MNRAS, 442, 2208 

\bibitem[Smith et~al.(2014)]{smith14}
Smith, R.~J., Glover, S.~C.~O., Clark, P.~C., Klessen, R.~S., \& Springel, V.\ 2014, MNRAS, 441, 1628

\bibitem[Sobolev(1957)]{Sobolev57} Sobolev, V.~V.\ 1957, SvA, 1678

\bibitem[Solomon et~al.(1987)]{sol87}
Solomon, P.~M., Rivolo, A.~R., Barrett, J., \& Yahil, A. 1987, ApJ, 319, 730

\bibitem[Springel(2005)]{springel05}
Springel, V.\ 2005, MNRAS, 364, 1105

\bibitem[Sternberg et~al.(2014)]{stern14}
Sternberg, A., {Le Petit}, F., Roueff, E., \& {Le Bourlot, J.}, 2014, ApJ, 790, 10

\bibitem[Strong \& Mattox(1996)]{sm96}
Strong, A.~W., Mattox, J.~R. 1996, A\&A, 308, L21

\bibitem[Visser, {van Dishoeck}, \& Black(2009)]{visser09}                                                         
Visser, R., {van Dishoeck}, E.~F., \& Black, J.~H.\ 2009, A\&A, 503, 323  

\bibitem[Ward, Wadsley \& Sills(2014)]{wws14}
Ward, R.~L., Wadsley, J., \& Sills, A.\ 2014, MNRAS, 439, 651

\bibitem[Wilson(1995)]{wilson95}
Wilson, C.~D.\ 1995, ApJ, 448, L97

\bibitem[Woodall et~al.(2007)]{umist07}
Woodall, J., Ag\'undez, M., Markwick-Kemper, A.~J., \& Millar, T.~J., 2007, A\&A, 466, 1197

\bibitem[Wolfire, Hollenbach \& McKee(2010)]{wolf10}
Wolfire, M.~G., Hollenbach, D., \& McKee, C.~F.\ 2010, ApJ, 716, 1191

\end{thebibliography}
\end{document}